\PassOptionsToPackage{unicode}{hyperref}
\PassOptionsToPackage{naturalnames}{hyperref}

\documentclass{aa} 

\usepackage[varg]{txfonts}
\usepackage{natbib}

\usepackage{graphicx}
\usepackage{txfonts}
\usepackage{hyperref} 

\hypersetup{
    colorlinks=true,
    citecolor=blue, 
    linkcolor=blue,
    filecolor=blue,      
    urlcolor=blue,
}

\usepackage{ulem}
\usepackage{amsmath}
\usepackage{amssymb}
\usepackage{placeins}  
\usepackage{subcaption} 
\usepackage{newtxtext,newtxmath}

\newcommand{\fourCothree}{4C\,+03.24}

\newcommand{\oiii}{{[\ion{O}{iii}]}}

\newcommand{\out}{\mathrm{out}}

\newcommand{\oii}{[\ion{O}{ii}]}
\newcommand{\oi}{[\ion{O}{i}]}
\newcommand{\cii}{[\ion{C}{ii}]}

\newcommand{\nii}{[\ion{N}{ii}]}

\newcommand{\neiii}{[\ion{Ne}{iii}]}
\newcommand{\sii}{[\ion{S}{ii}]}

\newcommand{\hei}{\ion{He}{i}}
\newcommand{\heii}{\ion{He}{ii}}

\newcommand{\lyalpha}{Ly$\,\alpha$}
\newcommand{\ha}{H$\,\alpha$}
\newcommand{\hb}{H$\,\beta$}
\newcommand{\halpha}{H$\,\alpha$}
\newcommand{\hbeta}{H$\,\beta$}
\newcommand{\hgamma}{H$\,\gamma$}
\newcommand{\hdelta}{H$\,\delta$}
\newcommand{\hepsilon}{H$\,\epsilon$}

\newcommand{\haa}{\mathrm{H\,\alpha}}
\newcommand{\hbb}{\mathrm{H\,\beta}}

\newcommand{\cmcubic}{$\rm{cm^{-3}}$}
\newcommand{\um}{{$\rm{\mu m}$}}
\newcommand{\mjy}{{$\rm{mJy}$}}

\newcommand{\mujybeam}{{$\rm{\mu Jy\,beam^{-1}}$}}
\newcommand{\kms}{{$\rm{km\,s^{-1}}$}}
\newcommand{\degree}{{$^{\circ}$}}
\newcommand{\WattHz}{{$\rm{W\,Hz^{-1}}$}}
\newcommand{\ergs}{$\rm{erg\,s^{-1}}$}
\newcommand{\ergscm}{$\rm{erg\,s^{-1}\,cm^{-2}}$}
\newcommand{\ergscmAA}{$\rm{erg\,s^{-1}\,cm^{-2}\,\si{\angstrom}}$}
\newcommand{\ergscmum}{$\rm{erg\,s^{-1}\,cm^{-2}\,{\mu}m}$}
\newcommand{\msun}{$\rm{M_{\odot}}$}
\newcommand{\lsun}{$\rm{L_{\odot}}$}
\newcommand{\msunyr}{$\rm{M_{\odot}\,yr^{-1}}$}

\newcommand{\mjykms}{$\rm{mJy\,km\,s^{-1}}$}

\usepackage{siunitx}
\DeclareSIUnit{\angstrom}{\textup{\AA}} 
\newcommand{\AAA}{\si{\angstrom}}

\newcommand{\logUcut}{$-3$}
\newcommand{\SNRflux}{2}
\newcommand{\SNRratio}{3}
\newcommand{\SNRratiomin}{1}

\newcommand{\neMin}{10}
\newcommand{\neMax}{10$^4$}
\newcommand{\RneMin}{0.4916}
\newcommand{\RneMax}{1.434}

\newcommand{\lagn}{$L_{\mathrm{AGN}}$}

\newcommand{\Mstar}{$M_{\mathrm{*}}$}

\newcommand{\mdot}{$\dot{M}_{\rm{out}}$}
\newcommand{\edot}{$\dot{E}_{\rm{out}}$}
\newcommand{\mdotpeak}{$\dot{M}_{\rm{out}}^{\rm{peak}}$}
\newcommand{\edotpeak}{$\dot{E}_{\rm{out}}^{\rm{peak}}$}
\newcommand{\mout}{$M_{\rm{out}}$}
\newcommand{\vout}{$v_{\rm{out}}$}
\newcommand{\iout}{$i_{\rm{out}}$}

\newcommand{\ebv}{$\rm{E(B\,{-}\,V)}$}
\newcommand{\neSII}{\ensuremath{n_{\rm{e,\sii}}}}
\newcommand{\Tee}{\ensuremath{T_{\rm{e}}}}
\newcommand{\npp}{\ensuremath{n_{\rm{p}}}}
\newcommand{\nee}{\ensuremath{n_{\rm{e}}}}
\newcommand{\neOut}{\ensuremath{n_{\rm{e,out}}}}
\newcommand{\neSIIcorr}{\ensuremath{n_{\rm{e,\sii}}^{\rm{corr}}}}
\newcommand{\neU}{\ensuremath{n_{\rm{e,U}}}}

\newcommand{\sfrA}{SF$_{\rm{A}}$}
\newcommand{\sfrB}{SF$_{\rm{B}}$}
\newcommand{\sfrC}{SF$_{\rm{C}}$}
\newcommand{\sfrEXT}{SF$_{\rm{ext}}$}

\newcommand{\colorOptical}{gray}
\newcommand{\colorUV}{blue}
\newcommand{\colorRadio}{green}

\newcommand{\colorXray}{orange}

\newcommand{\colorCIIa}{blue}
\newcommand{\colorCIIb}{yellow}
\newcommand{\colorCIIc}{orange}

\newcommand{\colorNucleus}{pink}

\newcommand{\vla}{VLA}
\newcommand{\jwst}{JWST}

\usepackage[dvipsnames]{xcolor}

\begin{document} 

    \title{
    The interplay between active galactic nucleus photoionization, radio jet, and star formation in the z\,$\sim$\,3.5 radio galaxy 4C\,+03.24
    }
    \titlerunning{AGN photoionization, radio jet, and star formation in 4C\,+03.24}
    \authorrunning{Dall'Agnol de Oliveira et al.}

\author{Bruno {Dall'Agnol de Oliveira}\inst{\ref{inst1}}
   \and Dominika {Wylezalek}\inst{\ref{inst1}}
   \and Pranav {Kukreti}\inst{\ref{inst1}}
   \and Wuji {Wang} \inst{\ref{inst2}}
   \and Rogério {Riffel}\inst{\ref{inst6}}
   \and Rogemar A. {Riffel}\inst{\ref{inst5}}
   \and Andrey {Vayner}\inst{\ref{inst9},\ref{inst2}}
   \and Caroline {Bertemes}\inst{\ref{inst1}}
   \and Julian T. {Groth}\inst{\ref{inst1},\ref{inst4}}
   \and David S. N. {Rupke}\inst{\ref{inst7}}
   \and Carlos {De Breuck}\inst{\ref{inst8}}
   \and Joël {Vernet}\inst{\ref{inst8}}
          }

   \institute{
Zentrum für Astronomie der Universität Heidelberg, Astronomisches Rechen-Institut, Mönchhofstr 12-14, D-69120 Heidelberg, Germany\label{inst1}\and
Caltech/IPAC, 1200 E. California Blvd. Pasadena, CA 91125, USA\label{inst2}\and
Departamento de Astronomia, Universidade Federal do Rio Grande do Sul, IF, CP 15051, 91501-970 Porto Alegre, RS, Brazil\label{inst6}\and
Departamento de F\'isica, CCNE, Universidade Federal de Santa Maria, 97105-900, Santa Maria, RS, Brazil\label{inst5}\and
Florida Gulf Coast University, 10501, FGCU Blvd. South, Fort Myers, FL 33965, USA\label{inst9}\and
School of Mathematics and Physics, The University of Queensland, St Lucia, QLD 4072, Australia\label{inst4}\and
Department of Physics, Rhodes College, 2000 North Parkway, Memphis, TN 38112, USA\label{inst7}\and
European Southern Observatory, Karl Schwarzschild Straße 2, 85748 Garching, Germany\label{inst8}
   }

   \date{Accepted: 17 September 2026}

 \abstract{
 High-redshift radio galaxies (HzRGs) are among the most powerful radio sources, and are associated with the most massive galaxies and dense environments at redshifts z\,$\gtrsim$\,1. They are ideal laboratories for studying how active galactic nucleus (AGN) events can shape the evolution of galaxies, as intense radiation, jets, and star formation can be observed simultaneously in these galaxies. We present JWST/NIRSpec integral field spectroscopy ($\sim$\,1.6\,kpc spatial resolution) of the 4C\,+03.24 system, a powerful HzRG at z\,$\sim$\,3.5 with a bolometric luminosity of $\sim$\,$10^{47.6}$\,{$\rm{erg\,s^{-1}}$}. We identified kinematically disturbed regions in the warm ($\sim$\,10$^4$\,K) ionized gas by decomposing the emission-line spectra into multiple Gaussian components, which is crucial to avoid overestimating the outflow properties. The outflow power peaks at $\sim$\,2\,kpc away from the nucleus, with a corresponding low kinetic coupling efficiency of $\sim$\,$8_{-5}^{+7}$\,{$\times$}\,10$^{-3}$\,\%. A combined analysis of the rest-frame optical and ultraviolet (from VLT/MUSE and HST imaging) continua revealed an extended emission (spanning $\sim$\,14\,kpc), which we interpret as partially tracing star-forming regions. With a clearly delineated bipolar morphology, we show that the AGN photoionization dominates the ionization of the interstellar medium along the radio jet axis. The {\cii}$\lambda$158{\um} emission gap in this region might be direct evidence of negative AGN feedback. We also discuss a possible scenario where 4C\,+03.24 could be situated in an overdense environment experiencing multiple galaxy interactions and the possibility of jet-induced star-formation.
}
   \maketitle
\nolinenumbers

\section{Introduction}\label{sec:intro}

Supermassive black holes (SMBHs) at the centers of galaxies play a crucial role in galaxy evolution \citep{heckman_best14}. As SMBHs grow, during active galactic nucleus (AGN) phases, enormous amounts of energy can be released and injected into the interstellar medium (ISM). The energy-ISM coupling can decrease ( ``negative feedback'') or increase (``positive feedback'') the star formation rate (SFR) in the host galaxy \citep{harrison_ramosAlmeida24}. 

The energy released by the AGN interacts with the ISM in different ways. This includes photoionization by high-energy photons \citep[e.g.,][]{liu+13a,storchi-bergmann+18} and the depletion of the molecular gas reservoir via photo or shock dissociation \citep[e.g.,][]{garcia-burillo+24,dallagnol+25a}, which could otherwise form new stars. The associated radiation pressure can also drive winds, observed as turbulence in the gas and/or outflows at different scales and in different phases of the gas \citep[e.g.,][]{cicone+18,harrison+18,ruschel+21,riffelRA+26}. These outflows can redistribute the gas in the host galaxy or even remove it in the most extreme scenarios. Outflows can also be driven by radio jets of relativistic plasma. 
As the jet propagates throughout the medium, it can inflate bubbles of hot gas and produce shocks, which can also ionize the gas \citep{mukherjee+18,mukherjee25}. As a consequence, one can observe multiphase gas outflows and/or turbulence injected into the gas in the disk \citep[e.g.,][]{riffel+14,venturi+21,riffelRA+23,deMellos+26}. Positive feedback can also happen with the shock compressing the gas and igniting local bursts of star formation \citep[SF; e.g.,][]{bicknell+00,gallagher+19,capetti+22,duggal+24}.   

These energetic phenomena are intermittent but have been observed in galaxies over a wide range of redshifts. In particular, at z\,$\sim$\,2, there is a peak in the volumetric density of the SMBH mass accretion rate and SFR \citep{madau_dickinson14}. Therefore, this is the epoch (1 <\,z\,< 3) when the SMBH and stellar masses in the galaxies accumulate faster,  commonly referred to as cosmic noon. This also seems to be the era when the kinetic output from both stars and AGN winds reaches a maximum, with the jet contribution peaking later, at z\,$\sim$\,1 \citep{heckman_best23}. Nonetheless, the most luminous radio galaxies are observed at higher redshifts. With volumetric densities peaking at 2\,$\lesssim$\,z\,$\lesssim$\,5 \citep[][]{jarvis+01, rigby+15,yuan+16}, these objects are called high-redshift radio galaxies (HzRGs) and are among the most powerful objects known. They are defined as sources with 500\,MHz rest-frame radio luminosities of $L_{500\,\rm{MHz}}$\,>\,$10^{27}$\,{\WattHz} and redshifts of z\,>\,2 \citep[as in][]{miley_deBreuck08}. 

High-redshift radio galaxies are associated with type 2 AGN sources, where nuclear radiation is obscured \citep{antonucci93,ramos-almeida_ricci17}, with the ionization axis oriented close to the plane of the sky. 
They are therefore well suited for mapping AGN feedback onto the ISM, as we can isolate the radiation emitted and absorbed by the gas and compare it with the stellar light from the host galaxies because there is low-to-no contamination from the point-like quasar emission. 
Additionally, we can disentangle the impact of radio jets, which can extend up to a few hundred kiloparsecs \citep[e.g.,][]{pentericci+00}, thus probing the connection to the circumgalactic medium. 

Often, HzRGs are dusty systems with rest-frame infrared (IR) luminosities of $L_{\rm{8\,-\,1000\,\mu m}}$\,$\gtrsim$\,10$^{12}$\,{\lsun} emitted by dust heated by AGNs and starbursts \citep{falkendal+19,drouart+14}. The dust, present in the torus and in the host galaxies, also helps collimate the radiation from the point source. 
In type 2 sources, this collimation can manifest itself as extended and asymmetrical photoionized regions with biconical morphology,  traced by emission lines such as {\oiii}$\lambda$5007 ({\oiii}, hereafter) \citep[e.g.,][]{schmitt+03,storchi-bergmann+18,fischer+18,venturi+21}. However, it can also scatter nuclear emission into our line of sight (LoS), ``contaminating'' observations in the rest-frame ultraviolet (UV) and optical regions. This is evidenced by the UV spectra of HzRGs being polarized by up to $\sim$\,20\,\% \citep{vernet+01}. 

Additionally, HzRGs are among the most massive systems and reside in some of the densest regions at their redshifts \citep{rocca-volrange+04,seymour+07,wylezalek+13,wylezalek+14}, with stellar masses of {\Mstar}\,$\sim$\,$10^{10.7}$\,--\,$10^{12.1}$\msun \citep{falkendal+19}. 
These sources have been used as signposts for proto-clusters \citep[e.g.,][]{rigby+14}, which are high-z dense environments that might become the most massive galaxy clusters in the local Universe. However, unlike local analogs, where the ``maintenance'' radio-mode feedback keeps the intergalactic medium hot (restraining further cooling), intense starburst events can be observed in these proto-clusters. Spectral energy distribution (SED) models of HzRGs indicate SFR\,$\sim$\,$10^1$\,--\,$10^3$\,\msunyr \citep{drouart+14,falkendal+19}, with a mean of $\sim$\,$110$\,\msunyr \citep{falkendal+19}, which hint that stellar feedback might be significant for radio galaxies at these redshifts. 
For z > 3 objects, the observed bright {\lyalpha} halos suggest that HzRGs are surrounded by large amounts of primordial hydrogen \citep[e.g.,][]{deBreuck+00}, and when compared to lower-z sources, UV emission line ratios indicate stronger starbursts \citep[]{villar-martin+07}. Therefore, despite being rare \citep{deBreuck+08,capetti+25}, HzRGs are key to understanding the evolution of massive galaxies, as they offer a view of their late-stage growth. 

With the advent of the James Webb Space Telescope (JWST), we can now study the impact of these phenomena in detail. With the integral field spectroscopic (IFS) mode of the Near Infrared Spectrograph (NIRSpec), the optical stellar and gas emission lines can be resolved at scales of $\sim$\,0.7\,--\,1.7\,kpc at 1\,<\,z\,<\,5 for typical full width at half maximum (FWHM) spatial resolutions of $\sim$\,0.1\,--\,0.2{\arcsec} \citep{vayner+24,deugenio+24}. 
We are now also able to resolve individual nearby companions and star-forming regions \citep{wang+25a,wang+25b,saxena+26}. 
Ionized outflows seem to be widespread in these powerful sources, hinting that AGN feedback might be crucial in the evolution of these proto-clusters, with jets and AGN photoionization-driven winds competing as the main driver of turbulence \citep{saxena+24,roy+26,solimano+25}.

In the present paper, we use JWST/NIRSpec observations to analyze rest-frame optical emission from {\fourCothree}, a source known for decades \citep{roettgering+94}.
This source has been studied in multiple wavelengths, from the radio \citep[e.g.,][]{vanOjik+95} to the X-ray \citep{smail+12}, due to being one of the most luminous HzRGs in multiple bands.  
Using this NIRSpec dataset, \citet{wang+25a,wang+25b} studied the properties of cloud companions identified by their kinematic deviations relative to the bulk gas. Here, we extend the analysis by performing a careful modeling of all optical emission lines. By isolating the emission from disturbed warm ionized gas ($\sim$\,10${^4}$\,K), we were able to spatially resolve the impact of the AGN feedback on this gas phase, and we study how it relates to radio and stellar emission.

\begin{figure}
    \includegraphics[width=1\linewidth]{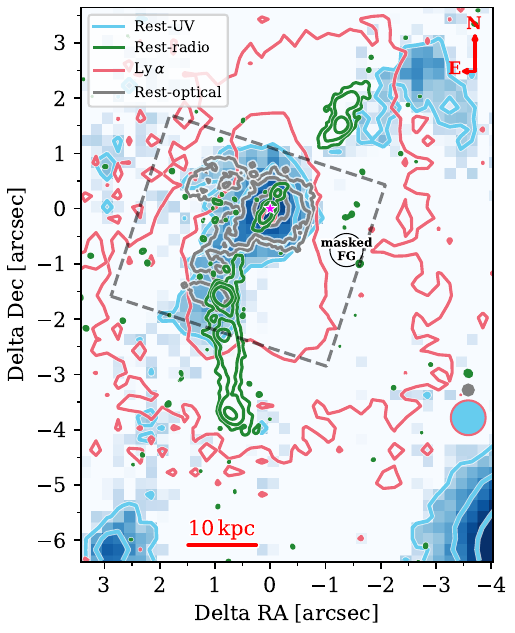}
    \caption{Overview of {\fourCothree} showing the emission distribution of the rest-frame VLT/MUSE UV continuum map. Overlaid are the contours from the rest-frame JWST/NIRspec optical (in {\colorOptical}) and the {\vla} 68\,GHz radio continua (in {\colorRadio}) along with the {\lyalpha} emission line total flux (in red). 
    The dashed line marks the NIRSpec FoV. North is up, and east is right, with the {\colorNucleus} star marking the AGN radio core nucleus. A nearby foreground (FG) object ($\sim$\,1.6{\arcsec} southwest of the nucleus) was masked in the rest-frame UV and optical images. Horizontal and vertical artifacts in the UV were also masked (see Fig.\,\ref{fig:muse-mask}). Contour levels are geometric progressions of the type of $\sigma_{\rm{rms}}$\,$\times$\,4$^{k}$, for $k$\,=\,[0,\,1,\,2,..] and with \,$\sigma_{\rm{rms}}$ values of 2.7\,{$\times$}\,$10^{-21}$\,{\ergscmAA} for the UV; \,1.1\,{$\times$}\,$10^{-19}$\,{\ergscmum} for the optical; 4.2\,{$\times$}\,$10^{-19}$\,{\ergscm} for {\lyalpha}; and 0.23\,{\mjy} for the radio. 
    The ellipses at the bottom right represent the FWHM spatial resolution of each image (color matched). 
    }
    \label{fig:fig0}
\end{figure}

The paper is organized as follows.
We first describe {\fourCothree} in detail in Sect.\,\ref{sec:object}, evoking previous works. In Sect.\,\ref{sec:data}, we present the JWST/NIRSpec observations; the auxiliary data used in the analysis is described in Appendix\,\ref{ap:data_ancillary}. The methodology used to derive gas properties is described in Sect.\,\ref{sec:analysis}. In Sect.\,\ref{sec:results}, we present and discuss the results, while conclusions are presented in Sect.\,\ref{sec:conclusions}. For this work, we assumed a systematic redshift of z\,$=$\,3.5657 \citep{nesvadba+17a}, corresponding to a relativistic systemic velocity of $v_{\rm{sys}}$\,=\,272346\,\kms. Assuming a $\mathrm{H_0=70\,km\,s^{-1}}$, $\mathrm{\Omega_M=0.3}$, and $\mathrm{\Omega_\Lambda=0.7}$ cosmology, this corresponds to an angular scale of 7.27 kpc/{\arcsec} and a luminosity distance of 31.26\,Gpc. We also assumed that the nucleus is localized at the {\vla} 15\,GHz radio core peak \citep{wang+25a}: 12h45m38.377s\,+03d23m21.14s (RA, Dec; J2000).

\begin{figure*}[!ht]
    \includegraphics[width=1\linewidth]{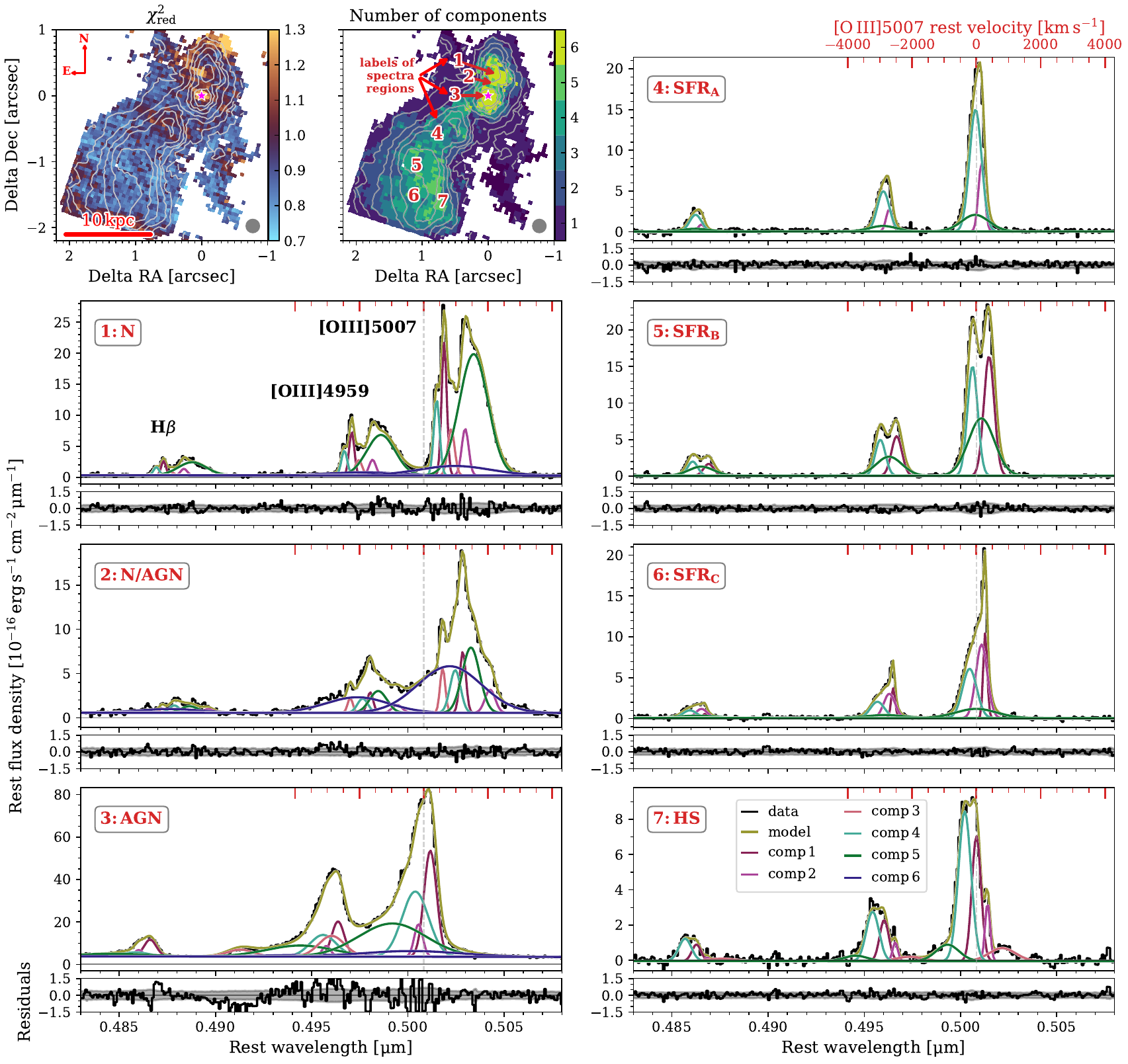}
    \caption{
    Example of the Gaussian models fitted to the data showing the spectral region covering the {\hb} and {\oiii} emission lines. In the upper left, maps of the reduced chi-square of the fit and the number of Gaussian components are displayed. White contours correspond to the {\oiii}$\lambda$5007 total flux map (as in Fig.\,\ref{fig:oiii_maps}). In the spectral panels, seven different examples are shown, and their label (e.g., ``1:\,N'') location is marked on the second map. In each spectral panel, the flux density of the data is plotted in black and the model in yellow, with each of its components represented with a different color. The top horizontal axes show the relative rest velocity of the {\oiii}$\lambda$5007 line, with the vertical dotted line marking 0\,{\kms}.}
    \label{fig:spec_fits}
\end{figure*}

\section{4C +03.24}\label{sec:object}

\object{4C\,+03.24} (also called 1243+036) was discovered and first identified as an HzRG due to its steep radio spectrum in a search for z > 2 radio galaxies \citep{roettgering+94}, and later spectroscopically confirmed to be at z\,$\sim$\,3.5 \citep{vanOjik+95}. 
In Fig.\,\ref{fig:fig0}, we present an overview of the system, with the datasets being described in Sect.\,\ref{sec:data} and Appendix\,\ref{ap:data_ancillary}.

With an integrated rest-frame 500\,MHz radio luminosity of $L_{\rm{500\,MHz}}$\,$\sim$\,$10^{29.2{\pm}0.1}$\,{\WattHz} \citep{falkendal+19}, {\fourCothree} is among the 5\,\% of the most luminous radio sources at 3\,<\,z\,<\,4 \citep{miley_deBreuck08}. This source was studied in detail by \citet{vanOjik+95}, combining 1.5\,--\,8.3\,GHz VLA radio images (0.2\,--\,1{\arcsec} resolution), with narrow-band imaging centered on {\lyalpha} and optical spectroscopy obtained with the ESO New Technology Telescope. Their work identified a prominent radio jet extending $\sim$\,1.85{\arcsec} ($\sim$\,13.5 \,kpc) to the northwest and southeast of the nucleus (see Fig.\,\ref{fig:fig0}). At this position in the southeast, there is a radio hotspot (HS) where the radio jet emission bends southward. In \citet{kukreti+26}, we show that there is a local enhancement in the gas disturbance in the HS, with the line flux ratios indicating a post-shock region to the south of this area.

Different authors have attempted to model the integrated SED \citep{cimatti+98,seymour+07,deBreuck+10,drouart+14,falkendal+19}. By combining photometry from 3.5\,{\um} to 6.5\,{GHz} (rest-frame), \citet{falkendal+19} isolated the synchrotron and dust emission (heated by the AGN and stars). From the stellar component, they estimated an SFR\,$\sim$\,$140_{-130}^{+240}$\,\msunyr. Following \citet{drouart+14}, we obtain an AGN bolometric luminosity of $L_{\rm{AGN}}$\,$=$\,6\,$\times$\,$L_{\rm{AGN,IR}}$\,$=$\,$10^{47.6\,\pm0.1}$\,{\ergs}, 
where $L_{\rm{AGN,IR}}$ is the AGN IR luminosity \citep{falkendal+19}. By combining old stellar population templates and thermal dust components in the rest-frame optical to mid-IR range, \citet{deBreuck+10} obtained an upper limit on the total stellar mass of $M_*$\,{$\lesssim$}\,$10^{11.3}$\,{\msun} for the system. \citet{kolwa+23} noted that later K-band photometry is consistent with the upper limits of \citet{deBreuck+10} models, suggesting that the $M_*$ should be very close to the published upper limit. 

For comparison, other HzRGs at 3 < z < 4 have $M_*$ ranges of {$\sim$}\,$10^{10.5}$\,--\,$10^{11.5}$\,{\msun} \citep{deBreuck+10}, SFR of $\sim$\,$80$\,--\,$630$\,{\msunyr} \citep{falkendal+19}, and $L_{\rm{AGN}}$ of $\sim$\,$10^{47.1}$\,--\,$10^{47.6}$\,{\ergs} \citep{falkendal+19}. When including type 1 sources, AGN can reach luminosities of {\lagn}\,$\sim$\,$10^{48}$\,{\ergs} \citep[see the compilation of][]{venturi+26}. Hence, although having an overall normal SFR and {\Mstar} for an HzRG at z\,$\sim$\,3.5, {\fourCothree} seems to host one of most powerful active nuclei at these redshifts.

{\fourCothree} has an extended ($\sim$\,250\,kpc) {\lyalpha} emission \citep{vanOjik+95,wang+23}. Due to its unusual bright emission, it is classified as a {\lyalpha}-excess object \citep[LAE;][]{villar-martin+07}. To explain such brightness and the flux ratios of UV lines, \citet{villar-martin+07} needed to combine AGN and stellar photoionization, discarding shocks, extremely high densities, and low metallicities as alternative scenarios. \citet{vanOjik+95} argued that the lower velocity dispersion and LoS velocities observed in the extended {\lyalpha} nebula could be tracing inflowing gas. Therefore, the associated gas may already have been located in the halo before the jet was launched, possibly originating from a previous galaxy merger. Using VLT/MUSE IFS observations, \citet{wang+23,ritter+26} modeled the {\lyalpha} profiles as a combination of an intrinsic Gaussian {\lyalpha} emission with multiple absorbers (assumed to be neutral hydrogen gas. \citet{wang+23} found an intrinsic (absorption-corrected) luminosity of $L_{\rm{Ly\alpha}}^{\rm{int}}$\,$\sim$\,$10^{45.46\pm0.02}$\,{\ergs}, a factor of $\sim$\,5 greater than the observed value. Furthermore, similar to other type 2 sources in their sample, {\fourCothree} shows a more asymmetric {\lyalpha} nebula, when compared to type 1 objects \citep{wang+23}. \citet{binette+98} analyzed the {\lyalpha} profile from a spectrum extracted at the radio hotspot. Contrary to the absorption scenario, they proposed that the observed multiple peaks in the line are caused by Fermi acceleration associated with jet-induced SF. 

Using Chandra observations, \citet{smail+12} reported a faint 0.5\,--\,8\,keV emission extending along the region covered by the radio jet. They argue that the spatial correlation between radio and X-ray can be explained by the Inverse Compton (IC) mechanism, in which X-ray emission arises from the interaction of the relativistic electrons in the jets with lower-energy radiation, particularly with the far-IR emission from dust-obscured starbursts. We note, however, that other authors claim that X-ray emission is consistent with IC from the Cosmic Microwave Background (CMB) radiation \citep{wu+17,hodges-kluck+21}. 

Different authors noted that the rest-frame UV-to-optical continuum of {\fourCothree} is co-spatial with the radio emission \citep{vanOjik+95,vanBreugel+98,pentericci+98}. This is a common feature in HzRGs \citep{chambers+87}, which might suggest jet-induced SF \citep{rees89}. Other scenarios include dust and/or gas scattering \citep{tadhunter+02}, nebular emission \citep{dickson+95}, stellar tidal debris from mergers \citep{pentericci+01}. 

The properties and kinematics of the optical emission lines were already studied using IFS from VLT/SINFONI \citet{nesvadba+17a,nesvadba+17b} and with our same JWST/NIRSpec dataset \citep{roy+26}. Similar to other HzRGs in their sample, \citet{nesvadba+17a} found solar to super-solar metallicities in the gas and that the {\oiii} kinematics is inconsistent with simple rotating systems. They also suggest that the AGN luminosity would be sufficient to unbind part of the gas, when considering the estimated stellar velocity dispersion of $\sigma_*$\,$\sim$\,300\,--\,350\,{\kms} \citep[based on their sample]{nesvadba+17b}. Similarly, \citet{roy+26} reported highly disturbed ionized gas, with the linewidth containing 80\,\% of the {\oiii} flux ($W_{80}$) reaching values of $\sim$\,2500\,{\kms} close to the nucleus. Both works support jets as the main drivers of the observed disturbance in the gas.

\section{{\jwst}/NIRSpec rest-frame optical data}\label{sec:data}

Near-IR integral field spectroscopic observations were carried out with the {\jwst}/NIRSpec instrument \citep{boker+22} during Cycle 1 (ID: JWST-GO-1970; PI: Wuji Wang), as originally described in \citet{wang+25a}. With a G235H/F170LP disperser/filter combination, the spectral region covers the wavelength range $\sim$\,1.7\,--\,3.1\,{\um} ($\sim$\,0.36\,--\,0.69\,{\um} rest-frame wavelength). Across the spectral region, the instrumental velocity resolution ranges from 34 to 68\,{\kms} (standard deviation). 
The FWHM spatial resolution is \,$\sim$\,0.2{\arcsec} ($\sim$\,1.6\,kpc) at 1.8\,{\um}, as measured in the foreground object continuum image (assuming that it is an unresolved point source).
The NIRSpec field of view (FoV) is highlighted in Fig.\,\ref{fig:fig0}, together with the rest-frame optical continuum map (see Sect.\,\ref{sec:fitting}, for more details).

The analysis and results shown in this paper were obtained after re-reducing the data cube. For this, we used the JWST Science Calibration pipeline, following the steps described for the first version of the data in \citet{wang+25a}, which is partially based on \citet{vayner+23}. Some additional steps were performed. Before merging the individual exposures, we created custom masks to account for micro-shutter assembly (MSA) leakage, which can introduce ghost features in the continuum if not carefully masked \citep[see][]{groth25}. In addition to sigma-clipping strongly illuminated spaxels, we masked bright artifact features in regions previously flagged in the data quality extension (DQ). Finally, after subtracting the background, we visually inspected different spectral channels in the merged data cube. We then manually masked bad pixels, further reducing spurious artifacts. We then modified the World Coordinate Systems (WCSs) keywords in the header to align the optical and the radio core peaks.

\section{Analysis}\label{sec:analysis}
\begin{figure*}[!ht]
    \centering
    \includegraphics[width=1\linewidth]{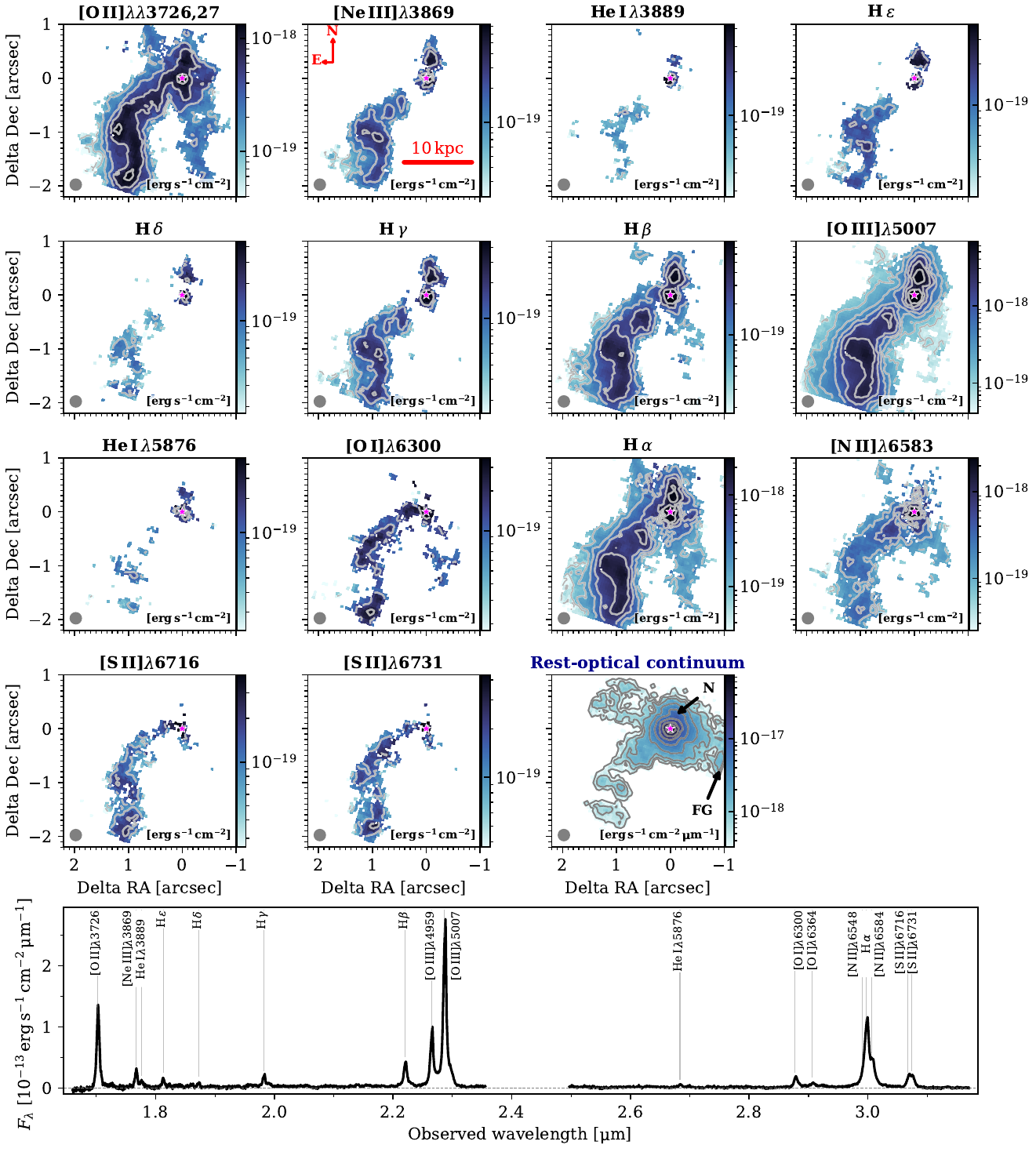}
    \caption{Maps of the total observed fluxes of the emission lines of {\fourCothree}. All fitted lines are shown except for the {\oiii}, {\nii}, and {\oi} doublets, where only one of the pairs is displayed. The bottom-right map shows the observed flux density distribution of the fitted NIRSpec rest-frame optical continuum, integrated in the 3968\,--\,6613\,{\AA} rest wavelength range. 
    Different from Fig.\,\ref{fig:fig0}, the continuum contour here shows increasing levels following 3\,$\sigma_{\rm{rms}}$\,$\times$\,2$^{k}$. 
    Over the continuum map, we highlight the position of the foreground object (FG) and of the north companion (``N''). 
    The {\colorNucleus} star marks the nucleus, as traced by the {\vla} 15\,GHz radio core peak.
    North is up, and east is right. 
    The bottom panel shows the integrated spectrum, using all spaxels where at least one emission line is detected. 
    }
    \label{fig:all_lines}
\end{figure*}

\subsection{Emission line fitting procedure}\label{sec:fitting}

We modeled the observed rest-frame optical continuum and the emission lines of the entire NIRSpec cube spaxel-by-spaxel. To characterize the continuum, we used a polynomial function. Each emission line was modeled with multiple Gaussian components, with each ``$i$''-component having an amplitude ($A_{i}$), a LoS velocity ($v_{i}$), and a velocity dispersion ($\sigma_{i}$, the standard deviation).  
The fitting procedure was performed using the Python software package \texttt{q3dfit} \citep{wylezalek+22,rupke+23}. A notable feature of this software is that it convolves the models with the instrumental dispersion. Therefore, $\sigma_{i}$ of the models corresponds to the intrinsic velocity dispersion of each component. 

For the fit, we assumed that the individual components in all different lines are kinematically tied: 
$v_{i}^{\hbb}$\,=\,$v_{i}^{\oiii}$\,=\,...\,=\,$v_{i}^{\rm{line}}$ and $\sigma_{i}^{\hbb}$\,=\,$\sigma_{i}^{\oiii}$\,=\,...\,=\,$\sigma_{i}^{\rm{line}}$, for each Gaussian component "$i$" and for all lines. We describe the fitting procedure below and refer the reader to Appendix\,\ref{ap:ap_fitting} for additional details.

(1) We first modeled the rest-frame optical continuum with a 6-degree polynomial using line-free spectral regions. The resulting model of each spaxel was then subtracted from the data. The next steps were applied to this continuum-subtracted cube.  

(2) To constrain the kinematics ($v_{i}$, $\sigma_i$) of each component ``$i$,''  we first fitted only the {\hb}\,+\,{\oiii}$\lambda\lambda$4959,5007 emission lines. The entire cube was modeled six times, each time with a different number of Gaussians (one to six). In other words, we obtained six different models, one for each total number of components. 
          
(3) To choose the best model but penalize for an increasing number of parameters, we used the Akaike information criterion \citep[AIC;][as defined in Appendix\,\ref{ap:ap_fitting}]{akaike74}. In each spaxel, we calculated the AIC for each one of the six different models. We then chose the simplest model (lowest number of components $N_g$) that satisfies the threshold criterion of $\Delta \rm{AIC}$\,= AIC$_{N_g+1}$\,$-$\,AIC$_{N_g}$\,<\,10 \citep[for a decisive evidence against the simpler model, as described in][]{kass_raftery95}. As a result, different spaxels have a different number of components. We label the resulting parameters of the model as $A_{i,\rm{AIC}}$, $v_{i,\rm{AIC}}$, and $\sigma_{i,\rm{AIC}}$. 

(4) Next, we fitted the entire cube again, including all emission lines. We fixed their kinematics to the result of the previous step, leaving only the amplitudes to vary: $v_{i}^{\rm{line}}$\,=\,$v_{i,\rm{AIC}}$ and $\sigma_{i}^{\rm{line}}$\,=\,$\sigma_{i,\rm{AIC}}$, for all lines and components $i$. The following emission lines were fitted together: 
{\oi}$\lambda\lambda$3726,29,
{\neiii}$\lambda$3869,
{\hei}\,$\lambda$3889,
{\hepsilon},
{\hdelta},
{\hgamma},
{\hbeta},
{\oiii}$\lambda\lambda$4959,5007,
{\hei}\,$\lambda$5876,
{\oi}$\lambda\lambda$6300,64,
{\nii}$\lambda\lambda$6548,83
{\halpha} and
{\sii}$\lambda\lambda$6716,31. 
The flux ratios of the line doublets were held fixed (see\,Appendix\,\ref{ap:ap_fitting}). 
Note that we modeled the {\oi}$\lambda\lambda$3726,29 pair as a single emission line. Only spaxels where the signal-to-noise ratio (S/N) of {\oiii}$\lambda$5007 amplitude ($A_i$, from the fitted profiles) were greater than three were fitted in this step.

(5) In some regions, particularly along the northeast and southwest of the nucleus, the {\oiii} emission is too weak to be used to constrain the gas kinematics. However, we could still detect other lines, such as {\halpha} and {\oi} (see the darker region in Fig.\,\ref{fig:ordering}). In these regions, the spectra of all lines were fitted together using a single Gaussian for each line.

(6) Together, the last two steps yielded maps of the fitted parameters ($A_{i}$, $v_{i}$, $\sigma_{i}$, for each one of the six Gaussian components ``$i$'') covering the entire FoV. In practice, however, these maps are ``unordered.'' For example, in the maps of component 1 ($A_{1}$, $v_{1}$, $\sigma_{1}$), neighboring spaxels do not trace a single ``physical structure'' (e.g., gas rotating in a disk). To mitigate the issue, we reorganized the components using the \texttt{COSMICube} software \citep{groth25}. The algorithm groups similar structures by minimizing spatial variance in the parameter maps ($A_{i}$, $v_{i}$, $\sigma_{i}$, for all 6 components). Further details about this algorithm, and how it was applied, are given in Appendix\,\ref{ap:ordering}. 

In Fig.\,\ref{fig:spec_fits}, we show examples of the best-fit model at different spaxels, focusing on the spectral region covering the {\hb} and {\oiii} lines. In the top left of the figure, we added the maps of the total number of Gaussian components. Six components were needed to model the spaxels close to the nucleus and in the north of it. They can vary substantially in these regions (see the spectra labeled as 1, 2, and 3 in the figure). Four or five components are also needed in regions up to $\sim$\,2{\arcsec} southeast of the nucleus. In the figure, we also display the map of the reduced chi-square of the fit: $\chi^2_{\rm{red}}$\,=\,$\chi^2/N_{\rm{free}}$, where $\chi^2$ is the chi-square (see Sect.\,\ref{ap:ap_fitting}), and $N_{\rm{free}}$ is the number of free parameters. Most of the spaxels display $\chi^2_{\rm{red}}$ values in the $\sim$\,0.8\,--\,1 range, although higher values are found in the nucleus and to the north of it. This is a consequence of the complexity of the data, which seems to require the number of Gaussian components to be even higher in these locations (see, for example, the residuals in the ``3-AGN'' model in Fig.\,\ref{fig:spec_fits}). 
Given the complexity of the model, we decided against including even more kinematic components.

\begin{figure*}
    \includegraphics[width=1\linewidth]{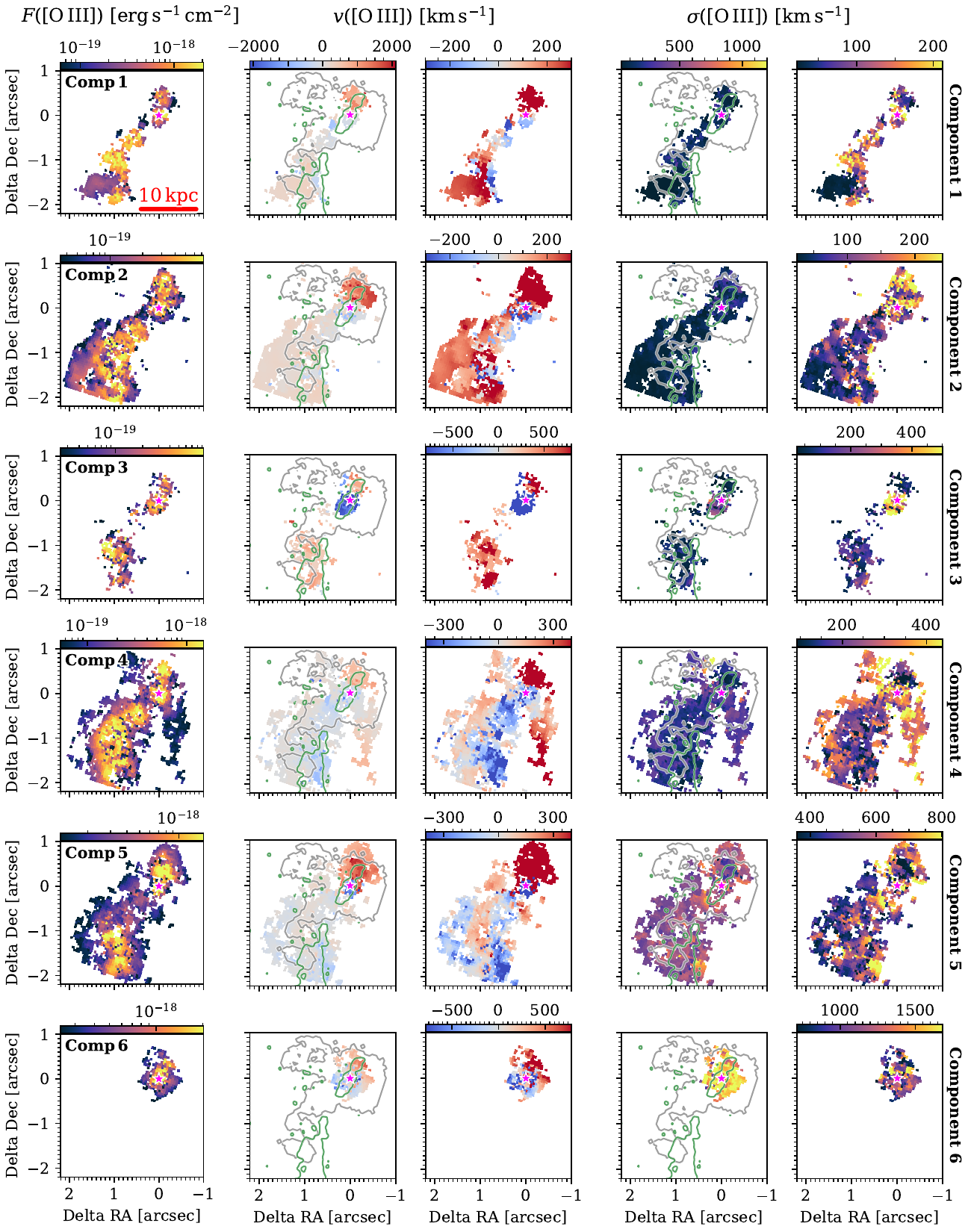}
    \caption{Maps of the Gaussian parameters of the six components fitted to the {\fourCothree} data.
    The dust-corrected flux is shown in column 1, the LoS velocity in columns 2 and 3, and the velocity dispersion in columns 4 and 5. The color bar ranges in columns 2 and 4 were fixed. The outer contours of the rest-frame VLA radio and the NIRSpec continua are shown in both columns 2 and 4.
    }
    \label{fig:oiii_maps}
\end{figure*}

In Fig.\,\ref{fig:all_lines}, we show the distribution of the total observed fluxes of all emission lines, except for doublets, where the flux ratio between them is fixed.  
The maps of the best-fit parameters of {\oiii}$\lambda$5007 are displayed in Fig.\,\ref{fig:oiii_maps}. The fluxes shown are already corrected for the dust attenuation (see Sect.\,\ref{sec:reddening}). To highlight the relative difference between components and the internal variation in each component, $v$ and  $\sigma$ maps are shown with two different color bar ranges. Only components with an S/N greater than {{\SNRflux}} are shown. The same threshold was used throughout the paper, except for quantities derived from flux ratios, where we used an S/N\,>\,{{\SNRratio}} threshold. We also note that the uncertainties in the Gaussian parameters of each fitted component were estimated using the unified equations from \citet{lenz_ayres92}.

\subsection{Dust attenuation}\label{sec:reddening}

To correct the observed fluxes for the effective dust attenuation effects, we used the observed ratio of the total fluxes of two Balmer emission lines ($\haa/\hbb$). From it, we derived the color excess ({\ebv}) in the nebular lines, as commonly done in the literature \citep[e.g.,][]{revalski+18}. We adopted the extinction curve of \citet{calzetti+00} with a total-to-selective extinction ratio of $R_v$\,=\,4.05. An intrinsic line ratio of $(\haa/\hbb)_0$\,=\,2.86 was assumed, characteristic of clouds with electron temperatures and densities of {\Tee}\,$\sim$\,$10^4$\,K and {\nee}\,$\sim$\,10$^2$\,--\,10$^3$\,{\cmcubic} \citep[for the case B of recombination]{osterbrock_ferland06}. 

We calculated the {\ebv} map from the smoothed {\ha}/{\hb} flux ratio, using a Gaussian kernel with FWHM\,=\,0.1{\arcsec} and following the steps described in Appendix\,\ref{ap:smoothing}.
Figure\,\ref{fig:ebv_ne} shows the smoothed {\ebv} map, while the non-smoothed version is displayed in Fig.\,\ref{fig:ebv_ne_raw}. 
The median {\ebv} value measured is $0.16_{-0.10}^{+0.16}$\,mag (uncertainty from the 15 and 84\,\% percentiles), and the mean is $0.20$\,mag. 
We note that before measuring the local effect in 4C\,+03.24, we corrected the fluxes for the Galactic dust attenuation. For this, we used NASA/IPAC Infrared (IRSA) tools to obtain attenuation maps from \citet{schlafly+11}, which uses the extinction law curve from \citet{fitzpatrick+99}. The mean color excess is E(B-V)$_{\rm{Gal}}$\,=\,0.0265. 

\subsection{Ionization parameter, BPT diagram}\label{sec:ionization_parameter}

To identify regions dominated by AGN photoionization, we analyzed how the flux ratio between different lines varies spatially, by making use of the ionization parameter and the Baldwin–Phillips–Terlevich (BPT) diagram \citep{bpt81}.  
Since we are interested only in the main ionization mechanism, we considered only the total fluxes of each emission line. We refer to \citet{kukreti+26} for an analysis of the ionization mechanism ``acting'' on components with different kinematic disturbances. 

\begin{figure*}[!ht]
     \centering
    \begin{subfigure}[b]{0.48\textwidth}
        \centering
    	\includegraphics[width=1\linewidth]{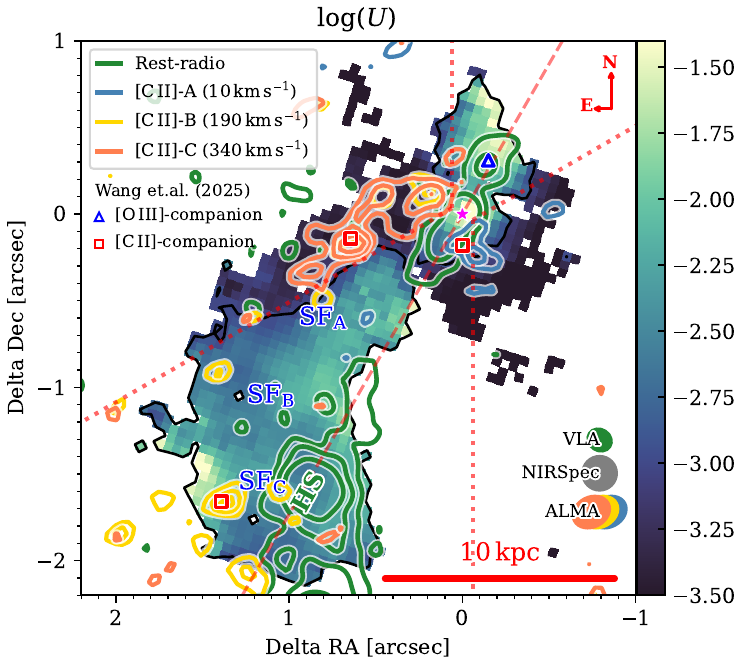}
    \end{subfigure}
    \begin{subfigure}[b]{0.48\textwidth}
        \centering
    	\includegraphics[width=1\linewidth]{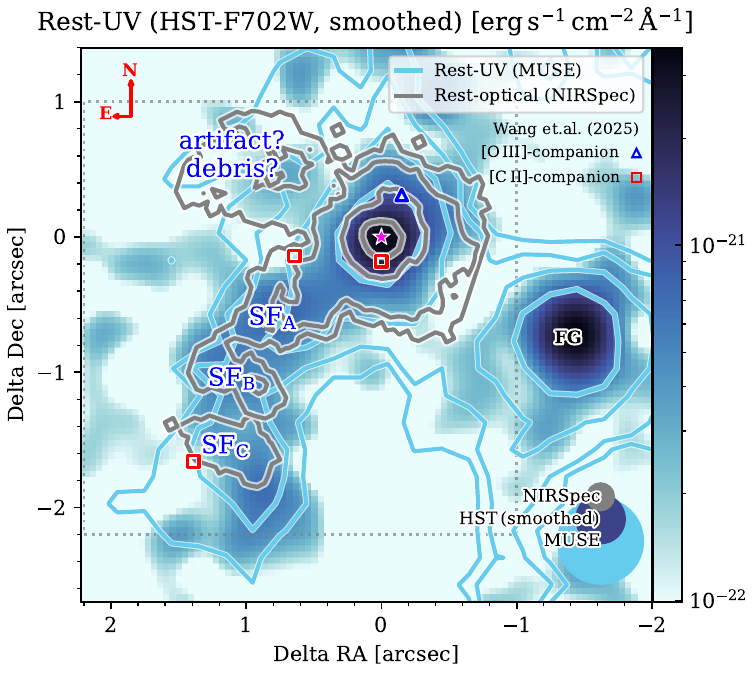}
    \end{subfigure}
        \caption{
        Left: 
            Logarithm of the ionization parameter ($U$), calculated from the {\oiii}/{\oii} (dust-corrected) flux ratio using a relation derived for local type 2 AGN \citep{carvalho+20}. The black contour corresponds to the level of $\log(U)$\,=\,{\logUcut}. The ionization axis (PA\,=\,-30{\degree}) is drawn in a dashed red line, with the biconical region (opening angles of 60{\degree}) delineated by dotted red lines. The {\colorCIIa}, {\colorCIIb}, and {\colorCIIc} contours correspond to different spectral channel cuts in the {\cii}$\lambda$158\,{\um} emission line, centered on different velocities \citep[as originally shown in][]{wang+25a}. Their levels follow [3,\,4,\,5..]\,$\times$\,$\sigma_{\rm{rms}}$ levels, with $\sigma_{\rm{rms}}$\,=\,0.44, 0.48, and 0.53\,{\mjykms} for \cii-A, \cii-B, and \cii-C, respectively.
        Right: 
            Flux density continuum distributions of {\fourCothree} showing the smoothed rest-frame UV image from the HST F702W band in the background (1300\,--\,1800\,\AA\ rest-frame wavelength range) 
            with contours of rest-frame MUSE UV (in {\colorUV}; 1029\,--\,2048\,\AA) and NIRSpec optical (in {\colorOptical}; 3968\,--\,6613\,\AA) continua (same as Fig.\,\ref{fig:fig0}) superposed on top. The artifacts present in the MUSE image were not masked here (see middle panel of Fig.\,\ref{fig:muse-mask}). The contour levels are the same as in Fig.\,\ref{fig:fig0}. The locations of the companions, as proposed in \citet{wang+25a}, are drawn as red squares. We mark in both maps the radio hotspot (HS), the three proposed extended star-forming regions ({\sfrA}, {\sfrB}, and {\sfrC}), and an area possibly affected by artifacts. The dotted square in the right panel corresponds to the FoV of the left panel.}
    \label{fig:logU_cont}
\end{figure*}

To gauge the radiation strength in different regions of ISM, we used the dimensionless ionization parameter ($U$). It is defined as the local ratio between the ionizing photon flux and the hydrogen density: 
\begin{equation}
    U = \frac{Q}{4\,r^2\,\pi n_{\rm{H}}\,c},
\end{equation}\label{eq:U}
where $c$ is the speed of light, $r$ is the radial distance from the ionizing source, $n_{\rm{H}}$ is the hydrogen density, and $Q$ is the rate of photons with sufficient energy (>\,$13.6$\,eV) to ionize the hydrogen. One can use the dust-corrected flux ratio between different Oxygen lines as a proxy of $U$. 
We used the equation derived by \citet{carvalho+20} for z\,<\,0.4 Seyfert galaxies using single-spectra measurements: 
$\log(U)\,{=}\,0.57\,{\times}\,[\log(\rm{O32)}]^2\,{+}\,1.38\,{\times}\,\log(\rm{O32})\,{-}\,3.14$, where O32 is the {\oiii}$\lambda$5007\,/\,({\oii}$\lambda$3726+3729) flux ratio.
The resulting log($U$) map of {\fourCothree} is shown in Fig.\,\ref{fig:logU_cont}.

\begin{figure}
    \includegraphics[width=1\linewidth]{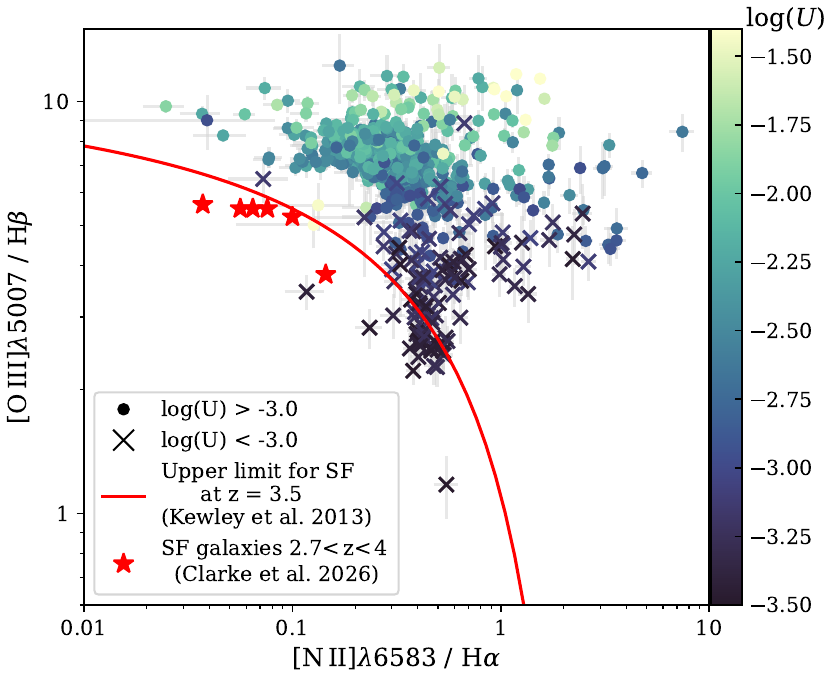}
    \caption{BPT diagram showing the distributions of the {\oiii}/{\hb} versus {\nii}/{\ha} flux ratios using the total flux of each line. Only spaxels in which all lines have an S/N (of the integrated profile) > 2 are shown, with one point per spaxel. The color represents the logarithm of the ionization parameter ($U$), with circle symbols for log($U$)\,>\,{\logUcut} and `x'  markers for log($U$)\,$\leq$\,{\logUcut}. The red curve marks the maximum line ratios expected for extreme stellar photoionization at z\,=\,3.5 \citep{kewley+13}. The red star corresponds to the mean integrated values obtained for SF galaxies at 2.7\,<\,z\,<\,4 for different stellar masses in the range of $10^{8.16}$\,--\,$10^{9.91}$\,\msun (from the left to the right) \citep{clarke+26}.
    }
    \label{fig:bpt}
\end{figure}

Additionally, in Fig.\,\ref{fig:bpt}, we display the BPT diagram of {\oiii}$\lambda$5007/{\hb} versus {\nii}$\lambda$6583/{\ha}. For reference, single-spectra observations of SF galaxies at 2.7\,<\,z\,<\,4 are included in the figure \citep{clarke+26}, from 35 sources covering {\Mstar}\,$\sim$\,$10^{8.2}$\,--\,$10^{9.9}$. We also show the line representing the maximum expected ratios for star-forming galaxies in extreme conditions at z\,=\,3.5 \citep{kewley+13}. These values are achieved only in star-forming regions with {\nee}\,$\sim$\,$10^3$\,{\cmcubic}, $\log(U)$\,>\,-2.9, and metallicities of 12\,+\,$\log(\rm{O/H})$\,<\,8.6.
Regions above this line are clearly due to AGN photoionization.

\begin{figure*}
    \includegraphics[width=1\linewidth]{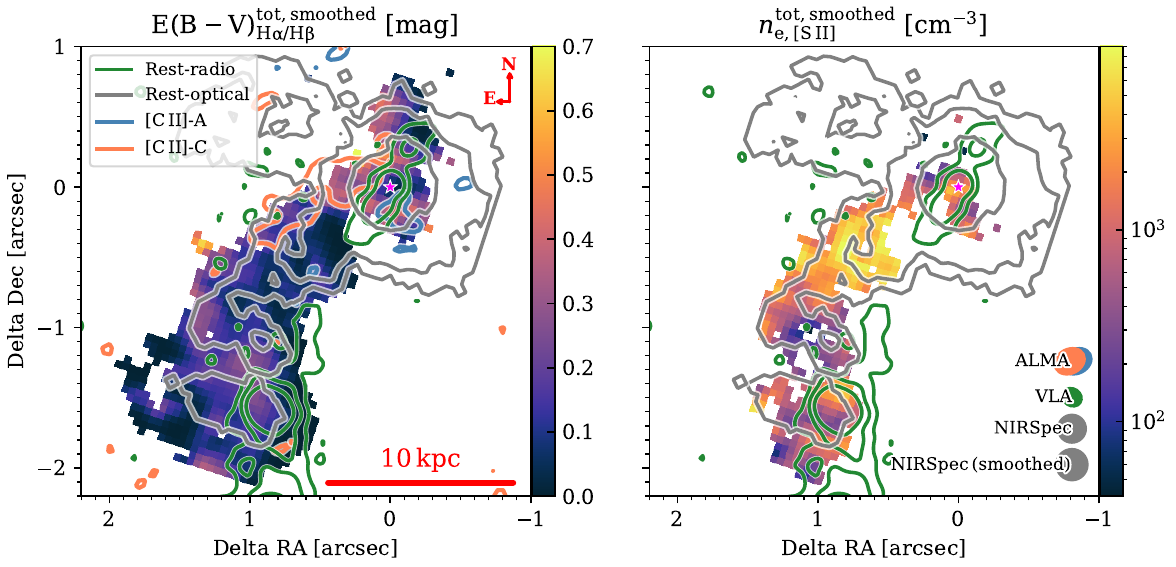}
    \caption{
        Smoothed color excess (left) and the electron density (right) of {\fourCothree}. Both quantities were calculated using the total fluxes of the emission lines. Only spaxels where all lines have S/N\,>\,{\SNRratio} are shown, with the remaining ones being filled with the median values of each parameter (see Appendix\,\ref{ap:smoothing}). The contours correspond to the rest-frame and optical continuum ({\colorOptical}) and radio ({\colorRadio}) flux distribution (as in Fig.\,\ref{fig:fig0}) and to the {\cii} channel cuts (only external level of {\cii}-A and {\cii}-C from Fig.\,\ref{fig:logU_cont}). In Fig.\,\ref{fig:ebv_ne_raw}, we show the non-smoothed version of these maps.}
    \label{fig:ebv_ne}
\end{figure*}

\subsection{Density}\label{sec:density}

Measurements of ionized gas mass depend on the electron density of the associated clouds. We first use the density diagnostic based on the {\sii}$\lambda$6716/{\sii}$\lambda$6731 total flux ratio \citep{sanders+16}, for a fixed electron temperature of {\Tee}\,=\,$10^4$\,K. In Fig.\,\ref{fig:ebv_ne}, we present the smoothed electron density (\neSII) map (see Fig.\,\ref{fig:ebv_ne_raw}, for the non-smoothed version). We applied the same smoothing method used for {\ebv} (see Appendix.\,\ref{ap:smoothing}), resulting in the median value of  {\neSII}\,$=$\,$1050_{-750}^{+1810}$\,{\cmcubic} (15 and 84\,\% percentiles). 

However, \neSII\ is not a good proxy for the density of the outflowing ionized gas (\neOut). 
As discussed by \citet{davies+20}, in a given photoionized cloud, the bulk of the emission from the {\sii} lines is produced in the partially ionized region, ahead of the ionization front, where most of the gas is neutral. However, most of the emission from the {\oiii} lines, which is a better tracer of the ionized outflow, is produced behind the ionization front. Therefore, the electron density where the {\oiii} and {\sii} emission mainly originates can be quite different, especially for higher densities. Corroborating this, {\neOut} values inferred from more reliable methods tend to be higher than {\neSII}, with some cases showing discrepancies of a few orders of magnitude \citep{davies+20,revalski+22,holden+26}. 
To account for such an effect in outflow-related measurements (see Sects.\,\ref{sec:mass} and \ref{sec:mdot_edot}), we applied the correction provided by \citet{holden+26}: $\log(\neSIIcorr)\,{=}\,\log(\neSII)\,{+}\,0.75\,(\pm0.07)$.  This relation was obtained by comparing the mean {\sii}-based values with the ones derived using trans-auroral lines \citep[method originally introduced by][]{holt+11}, which are more sensitive to higher densities. The comparison uses the total fluxes of the lines, given that the effect of using a broad (outflowing) component was negligible in their analysis. Note that it was derived for local type 2 AGN, and we are extending it to an object at z\,$\sim$\,3.5. Additionally, we assumed that this correction remains valid for spatially resolved measurements, although it was derived from single-aperture spectra.

We can also use the ionization parameter to estimate the electron density \citep{baron+19}. Using equation \ref{eq:U} and assuming a fully ionized gas ({\nee}\,$\sim$\,$n_H$), we get an electron density radial profile of: {\neU}\,=\,$Q\,/(4\,r^2\,\pi U\,c)$. We followed \citet{baron+19} and used the {\fourCothree} AGN luminosity \lagn\,$=$\,$10^{47.6\,\pm0.1}$\,{\ergs} \citep{falkendal+19} to obtain the ionizing photon rate: 
$Q$\,$=$\,10$^{10.06}$\,$\rm{s^{-1}}\,{\times}$\,({\lagn/{\ergs}})\,$\sim$\,$10^{57.7}$\,$\rm{s^{-1}}$. 
This is based on their SED model for a spinless black hole with a mass of $10^7$\,{\msun}, with the corresponding mean energy of ionizing photons at 36\,eV. As noted by \citet{baron+19}, there is a 10\,\% difference in $Q$ depending on the assumed SED. 
Since we are assuming an AGN ionizing source, we measured {\neU} only in regions dominated by AGN photoionization (log($U$)\,>\,{\logUcut}, as highlighted in the left panel of Fig.\,\ref{fig:logU_cont}).

\subsection{Ionized gas mass and outflow velocity}\label{sec:mass}

Following \citet{cano-diaz+12}, we derived the mass of the disturbed ionized gas from the {\oiii}$\lambda$5007 dust-corrected luminosities (L$_{{\oiii}}$), using the equation 
\begin{equation}
    M_{\rm{out}} = \frac{5.33\,{\times}\,10^7\,{\rm{M_{\odot}}}}{10^{\rm{[O/H]}}}\,\left[\frac{L_{{\oiii}}}{10^{44}\,\rm{erg\,s^{-1}}} \right] \left[ \frac{10^3\,\rm{cm^{-3}}}{n_{\rm{e,out}}} \right],
    \end{equation}
where $n_{\rm{e,out}}$ is the electron density of the outflowing gas, and [O/H] is the oxygen abundance relative to the solar value. 
The above equation assumes that the majority of the oxygen is double-ionized, and a uniform density for all ionized gas clouds covered by a given spaxel. For our calculations, we assumed a fixed solar metallicity. However, as discussed in Appendix\,\ref{sec:metallicities},  $Z$\,$\sim$\,1\,$Z_{\odot}$ values are found only around the nucleus, with lower values elsewhere.
By using the derived $Z$ map, local values of gas mass would change by factors of $\sim$\,$1/10^{\rm{[O/H]}}$\,$\sim$\,$1/(Z/Z_{\odot})$\,$\sim$\,0.9\,--\,5, with the larger differences being observed at the north of the nucleus and close to the radio hotspot.

In the literature, different definitions have been used to characterize the outflow velocity \citep[see][for an example of the effect of using different velocity definitions]{dallagnol+21}. 
Here, similar to \citet{rupke+13}, we define the outflow velocity as 
\begin{equation}
    v_{\rm{out}} = |v_{\rm{gas}} - v_{\rm{rest}}| + 2\,\sigma_{\rm{gas}},
\end{equation}
where $v_{\rm{gas}}$ and $\sigma_{\rm{gas}}$ are the LoS velocity and standard deviation of a given component. In addition, $v_{\rm{rest}}$ is the LoS velocity that the ionized gas would have if it were not disturbed by AGN feedback (e.g., the rotation velocity of clouds moving in an orderly manner in a disk). We set $v_{\rm{rest}}\,{=}\,v_{\rm{sys}}$ everywhere and for all components, ignoring rotation and the possibility of clouds being launched from a companion (with a different $v_{\rm{sys}}$).

\subsection{Mass outflow rate and outflow power}\label{sec:mdot_edot}

The mass outflow rate (\mdot) and outflow kinetic power (\edot) were measured in semi-annuli circular regions, with a fixed $\Delta r$ projected width. Below, we use $r$ and $v$, for projected radii and LoS velocities, respectively, with $R$ and $V$ referring to their de-projected values. At each annulus, we calculated the values of these quantities using
\begin{eqnarray}
    \dot{M}_{\rm{out}}(R) = \frac{M_{\rm{out}}(R)\,V_{\out}(R)}{\Delta R}\\
    \dot{E}_{\rm{out}}(R) = \frac{1}{2}\dot{M}_{\rm{out}}(R)\,V^2_{\out}(R).
\end{eqnarray}
In the equations, \mout$(R)$ is the sum of the warm ionized mass from the outflowing material, and \vout$(R)$ is its average outflow velocity at the radius $R$. Both quantities are measured inside each semi-annulus. No correction for projection effects was applied, meaning that $V\,{=}\,v$ and $R\,{=}\,r$. 
The semi-annuli are aligned with the ionization axis and have a fixed de-projected width of ${\Delta R}$\,=\,1\,kpc. 
In Sect.\,\ref{sec:density_outflow}, we tested the effect of assuming outflow inclinations (relative to the plane of the sky) of {\iout}\,=\,10 and 20{\degree}, using the corrections: $V\,{=}\,v\,{/}\,\sin(i_{\rm{out}})$ and $R\,{=}\,r\,{/}\,\cos(i_{\rm{out}})$.

To avoid overestimating  \mdot\ and \edot\, only Gaussian components that trace the disturbed ionized gas should be used in the calculations. However, as discussed in \citet{venturi+26}, different methods have been proposed to isolate them \citep[e.g.,][]{mullaney+13,zakamska_greene14,dallagnol+21}. In this work, for the {\mdot} and {\edot} measurements, we considered all Gaussian components with $\sigma_{\rm{gas}}$\,>\,400\,{\kms} as outflowing gas. This cutoff was chosen to isolate clouds with kinematics that cannot be explained by gravitational motion based on the dynamical arguments of \citet{nesvadba+17b}, which yields a $\sigma_*$\,$\sim$\,300\,--\,350\,{\kms} for their HzRG sample (which {\fourCothree} is part of). Note that they did not measure $\sigma_*$ directly, but inferred it from the dynamical mass. 
We also experimented using a more relaxed threshold of 200\,{\kms} (see Sect.\,\ref{sec:density_outflow}).

\section{Results and discussion}\label{sec:results}

The maps of the best-fit Gaussian parameters for the {\oiii}$\lambda$5007 are displayed in Fig.\,\ref{fig:oiii_maps}. The decomposition into multiple components is important to isolate the disturbed gas content. In \citet{kukreti+26}, we used such identification to analyze the ionization mechanism of the gas with disturbed kinematics. Additionally, we have applied the COSMICube \citep{groth25} algorithm to organize the parameter maps, which facilitates the interpretation of the physical condition of the gas traced by each Gaussian component. Whenever important, we mention individual components in the analysis below. 
However, the correct ordering of the component does not influence the main results of this work.  
This happens because we only considered total flux measurements for measurements derived from flux ratios (e.g., ionization parameter), while the outflow kinematics are based on velocity dispersion cuts that do not depend on the correct ordering of the components. 

\subsection{AGN photoionization marking a biconical region}\label{sec:agn_photo-ionization}

As shown in see Fig.\,\ref{fig:logU_cont}, the ionization parameter maps display high values close to the radio jet axis, reaching $\log(U)$\,$\sim$\,$-1.5$, with values as low as $\sim$\,$-3.5$ perpendicular to it.
High $U$ values indicate that an energetic ionization source, such as an AGN, is responsible for ionization of the clouds. For comparison, in local star-forming galaxies, models usually require $\log(U)$ values below $\lesssim$\,$-2.5$ \citep{dopita+00}. However, high $U$ values can also be observed in non-AGN regions, especially at high redshifts. For example, star forming galaxies at z\,$\sim$\,2.7\,--\,6.3 commonly show $-3$\,$\lesssim$\,log($U$)\,$\lesssim$\,$-2$, reaching values of $\sim$\,$-1.5$ \citep{reddy+23}. 

Nonetheless, the ionization parameter map shows a quite distinct biconical morphology (see Fig.\,\ref{fig:logU_cont}), typical of type 2 AGN \citep[e.g.,][]{wilson+93,storchi-bergmann+18,venturi+21,dallagnol+23}. We define here the $\log(U)$\,>\,{\logUcut} line as a region dominated by the AGN photoionization, which is confirmed by their respective ratio being clearly above the SF region in the BPT diagram (see Fig.\,\ref{fig:bpt}). We also defined a projected biconical region that marks the AGN ionization bicone (dotted lines in Fig.\,\ref{fig:logU_cont}). It was constructed assuming that the projected AGN ionization axis lies along PA\,=\,-30{\degree}, which is the direction of the inner radio jet, observed close to the nucleus ($\lesssim$\,0.5{\arcsec}). We also assumed a symmetrical opening angle of 60{\degree}, a value that matches the $\log(U)$\,=\,{\logUcut} contour in the southeast, and approximately covers the same contours in the north.

Therefore, based on Figs.\,\ref{fig:logU_cont} and \ref{fig:bpt}, the biconical region traces gas with flux ratios that cannot be explained by stellar ionization alone. Since this is a luminous radio galaxy, we cannot fully discard the contribution from shock ionization from jets (and from winds). \citet{kukreti+26} show that shocks are the main ionizing mechanisms in the region covered with $\log(U)$\,<\,{\logUcut} in {\fourCothree}, and in the post-deflection area at the south of the radio hotspot. Nonetheless, in general, AGN radiation is the main mechanism for ionizing the gas in the {\fourCothree} system. This is consistent with previous works, which found that the gas is predominantly AGN photoionized in HzRGs  \citep{vernet+01,humphrey+06}. This does not mean that shock-ionization does not happen inside the bicone, just that it is ``outshined'' by the AGN radiation \citep{venturi+21,riffelRA+21}. The same is true for stellar photoionization, which should be important but non-dominant in the nucleus and in extended regions in the southeast of it (see Sect.\,\ref{sec:sfr_ext}).

\subsubsection{{\cii} relation with the dust attenuation and the AGN ionizing bicone}\label{sec:cii}

In Fig.\,\ref{fig:logU_cont}, we added contours showing the flux distribution of the {\cii}$\lambda$158\,{\um} emission line, centered on three different velocities. 
The {\cii} line is associated with photodissociation regions, and has been used as a tracer of cold gas in high-redshift galaxies \citep{lagache+18}.
Interestingly, the emission close to the nucleus ({\cii-A} and {\cii-B}) shows a peculiar distribution. It seems to be restricted to the edges of the AGN ionizing bicone, suggesting that the cold gas might have been depleted by the strong AGN radiation. If so, this could be interpreted as direct evidence of negative AGN feedback on the ISM, at least temporarily, since these clouds could otherwise collapse and form new stars.
Additionally, the {\cii} distribution is spatially correlated with regions with higher dust attenuation (see Fig.\,\ref{fig:ebv_ne}). 
The smoothed color excess map displays values of $\sim$\,0.3\,--\,0.7\,mag around the nucleus and close to the {\cii} location,    
and above the median value of $0.16_{0.10}^{0.18}$\,mag. 

Close to the nucleus, the small-scale radio jet ($\lesssim$\,0.5{\arcsec}) likely plays a role in destroying {\cii}. In favor of this, we observed that there are two local radio peaks -- one at the nucleus and the other at $\sim$\,0.3{\arcsec} to the north-west of it -- that partly ``fill'' the nuclear area devoid of {\cii} emission. In these regions, we also observe low attenuation ({\ebv}\,$\sim$\,0\,mag in the smoothed map). 
However, note that, on the non-smoothed map (Fig.\,\ref{fig:ebv_ne_raw}), some of these spaxels have {\ha}\,/{\hb}\,<\,2.86 values (minimum value assumed for case B, as explained in Appendix\,\ref{ap:smoothing}).
These are also the regions with the most complex emission line profiles, with our best-fit models requiring 6 Gaussian components, which calls for caution in the interpretation. 
Therefore, given that {\ha} is quite blended with {\nii} in these spaxels -- increasing the uncertainty in total {\ha} flux -- we cannot rule out that the {\ebv} might have been underestimated locally.

\subsection{Extended rest-frame UV and optical emission}\label{sec:continuum}

It has long been known that the flux distribution of the rest-frame UV \citep{roettgering+95,pentericci+98,wang+25a} and optical \citep{vanBreugel+98} of {\fourCothree} is extended, peaking in the nucleus and stretching along the south-southeast of it. In these works, it was also noted that the continuum appears to be spatially correlated with the ionized gas and radio jet emission. With the high quality of the NIRSpec observations, we are now able to spatially resolve the rest-frame optical continuum at scales of $\sim$\,0.2{\arcsec} ($\sim$\,1.6\,kpc). Furthermore, the polynomial fitting of the continuum allowed us to remove the contribution from the ionized gas (see map in Fig.\,\ref{fig:all_lines}). 

\begin{figure}
    \includegraphics[width=.98\linewidth]{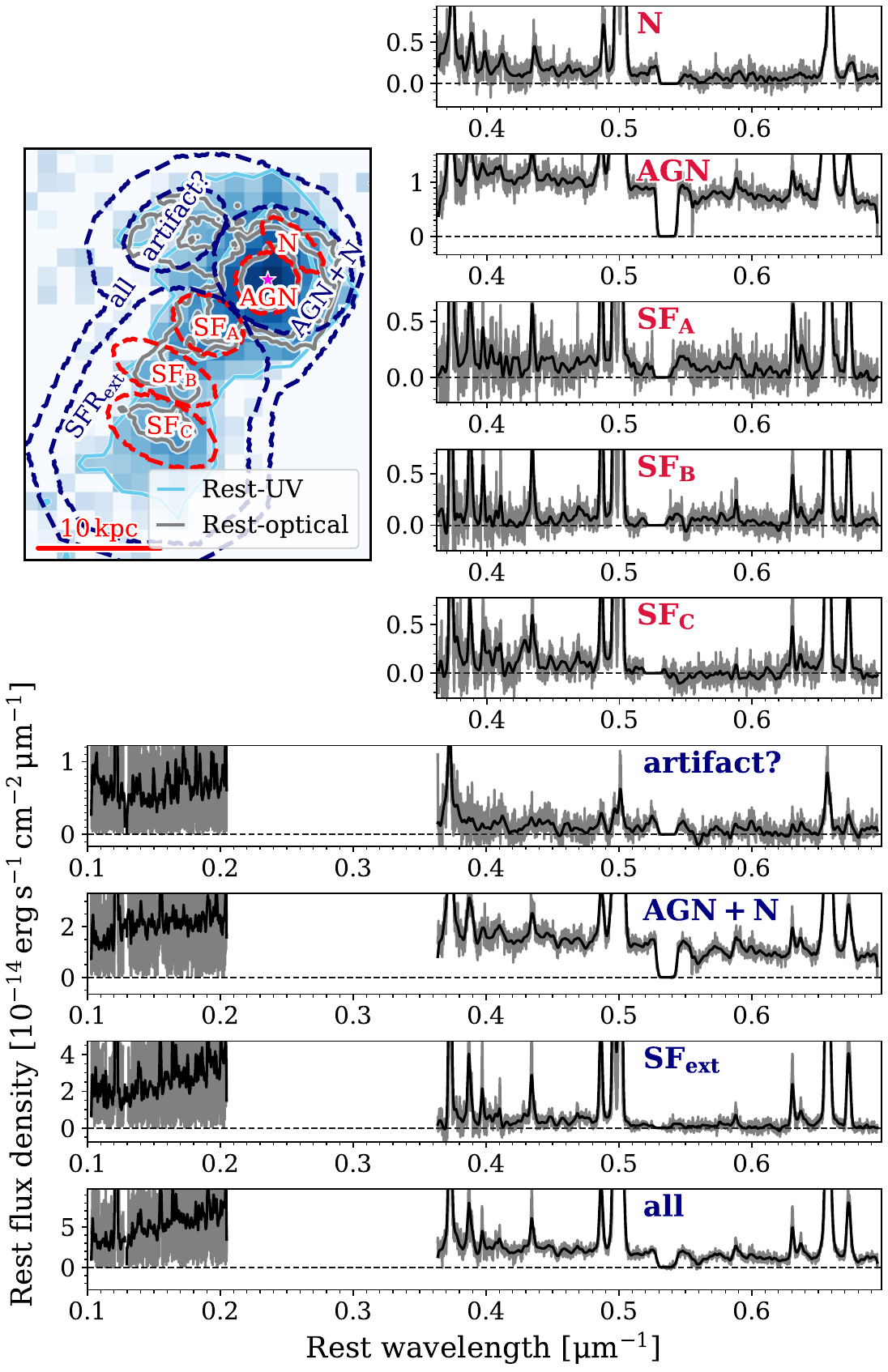}
    \caption{Integrated spectra from different regions in the {\fourCothree} system. 
    The regions are marked over the MUSE rest-frame UV image in the inset axis. 
    The NIRSpec integrated spectra are shown for all rows, while the MUSE spectra are shown only for the last four rows (regions). 
    For these, the rest-frame optical spectra were obtained after matching the NIRSpec spatial resolution with the MUSE one (see Appendix\,\ref{ap:spectra_region}). 
    For better visualization, we show the integrated spectra in gray together with a further smoothed version in black (using a 10\,pixel-$\sigma$ Gaussian kernel).
    }
    \label{fig:spectra_regions}
\end{figure}

In this paper, we tested a scenario where the bulk of extended rest-frame UV and optical continua are tracing young SF \citep[$\lesssim$\,200\,Myr time scales,][]{kennicutt_evans12}. For this, we considered that the rest-frame optical and UV continua have the same origin. More specifically, we used the higher spatial resolution in the rest-frame optical to better localize the SF regions, with the UV being used to confirm that it is not an artifact. In Fig.\,\ref{fig:logU_cont} (right panel), we display the map of the rest-frame HST UV continuum, overlaid by the contour images of the rest-frame VLT/MUSE UV and the JWST/NIRSpec optical continua, where we can identify extended emission. In addition, we also integrated the continuum in different regions of interest, as displayed in Fig.\,\ref{fig:spectra_regions}. Below, we describe and discuss these regions.

We first note that there is a strong UV peak at $\sim$\,1.6{\arcsec} to the southwest of the nucleus, which is consistent with a featureless continuum from a foreground object \citep{nesvadba+17a,wang+25a}. We exclude it from our analysis.

\subsubsection{Contribution from scattered light and nebular emission}\label{sec:scattering_nebular}

Although commonly used as an indicator of recent SF, the UV continuum can also be produced by other phenomena, such as nebular continuum emission and scattered light. Here, we discuss these two scenarios. 

We first note that, using spectropolarimetry observations, \citet{vernet+01} showed that 11$\pm$4\,\% of the integrated rest-frame UV continuum (1250\,--\,1400\,{\AA}) from {\fourCothree} is polarized, which comes from part of nuclear radiation that is scattered toward us. Although the continuum in both bands peaks at the {\fourCothree} nucleus, we cannot be sure if the polarization arises from this region. This happens because the observation was taken with a PA\,=\,156{\degree} slit angle \citep{vernet+01}. Hence, the slit also covers the extended emission in the southeast. 

In addition, based on the flux of the {\heii}\,$\lambda$\,1640 emission line, \citet{vernet+01} argue that nebular emission contributes to only $\sim$\,3\,\% of the rest-frame UV continuum. Using an analogous method (see Appendix\,\ref{ap:spectra_region}), we calculated the expected nebular continuum based on the {\hbeta} flux, with the emissivities obtained from the \texttt{Pyneb} Python package \citep{pyneb_15}. Overall, we find that the nebular emission -- inside the 1400\,--\,2000\,{\AA} rest wavelength range -- contributes to: 5$\pm$2\,\% around the continuum peak (``quasar+N'' in Fig.\,\ref{fig:spectra_regions});  9$\pm$4\,\% in the extended region in the southeast (``{\sfrEXT}'');  0.8$\pm$0.4\,\% in the east (``artifact'');  and 7$\pm$3\,\% in the entire region (``all''). In the rest-frame optical (3800\,--\,4700\,{\AA}), we found values of 2.2$\pm$0.2\,\% (``quasar+N''), 14$\pm$5\,\% (``{\sfrEXT}''),  0.9$\pm$0.3\,\% (``artifact''),  and 5$\pm$1\,\%  (``all'').

We proceed with the analysis assuming that the nebular continuum and scattered light are likely insufficient to explain the observed extended continuum emission. In other words, we propose that at least part of this emission comes from massive young stars. 
However, only new spatially resolved polarimetry observations could discard a higher percentage of scattered light contamination in the extended continuum emission.  
We note that nonnuclear UV polarization has been observed in some sources, with the peak of polarization detected up to a few kiloparsecs away from the nucleus, for both local and higher redshift sources  \citep[e.g.,][]{barnouin+23,barnouin+25,assef+25}.

\subsubsection{Extended star-forming regions}\label{sec:sfr_ext}

After inspecting the rest-frame optical continuum flux distribution outside the nucleus, we propose that it traces local ongoing SF in three different regions. They are observed along the southeast direction of the nucleus, separated from it by $\sim$\,1.0, 1.5, and 1.9{\arcsec}, which we named {\sfrA}, {\sfrB}, and {\sfrC}, respectively (see Fig.\,\ref{fig:logU_cont}). Collectively, we call these extended star-forming regions {\sfrEXT}, which we define as star-forming regions not directly associated with the {\fourCothree} host. In this definition, the {\sfrEXT} regions include companion galaxies and tidal features of ongoing interactions (from the main galaxy or the interacting object). Note that, due to its relative proximity to the nucleus ($\sim$\,7\,kpc distance), the {\sfrA} region could also be considered as a part of the central galaxy hosting the AGN. In all three regions, the rest-frame UV continuum is also detected, reinforcing the SF scenario since the radiation from young stars peaks at smaller wavelengths. This is highlighted by the integrated spectra shown in Fig.\,\ref{fig:spectra_regions}, where we observe a rise in flux density at lower wavelengths in the optical, with a subsequent decline in the UV (for {\sfrEXT}), possibly due to dust attenuation.

There is also a weak loop-shaped feature in the rest-frame optical emission, at the northeast of the nucleus, extending up to a $\sim$\,1.4{\arcsec} radius (labeled ``artifact?'' in Fig.\,\ref{fig:logU_cont}). It coincides with an elongation in the MUSE rest-frame UV continuum. In this region, {\oii} and {\ha} are the strongest emission lines (Fig.\,\ref{fig:all_lines}). These two lines have been used as indicators of SF \citep[e.g.,][]{calzetti+04}, covering time scales of $\lesssim$\,10\,Myr \citep{kennicutt_evans12}. Nonetheless, the gas distributed in this region is dominated by shock-ionization \citep[see][]{kukreti+26}, although this does not exclude the presence of SF. Additionally, the elongated MUSE UV emission is near the region masked due to mosaicking artifacts (see Fig.\,\ref{fig:muse-mask}). The emission is also not detected in the smoothed HST rest-frame UV image, although this could be due to the observation not being deep enough. Therefore, we cannot fully discard the possibility that the observed ``loop'' is an artifact. 

We further discuss evidence of companions in Appendix\,\ref{sec:metallicities}. There, we use a derived gas metallicity ($Z$) map (see Fig.\,\ref{fig:Z_Z0}) to discuss broad differences in $Z$ along the FoV. 
We used additional information taken from the {\oiii} kinematics and the {\cii} flux distribution to further discuss evidence of companions. Broadly, the $Z$ map indicates that {\fourCothree} has $Z$\,$\sim$\,$Z_{\odot}$ gas close to the nucleus, while being surrounded by more metal-poor content. This suggests that the main galaxy is undergoing multiple interactions with objects containing more pristine gas. 
However, we emphasize that these metallicity measurements are quite uncertain, as discussed in the appendix.

As detailed in Appendix\,\ref{ap:sfr}, we estimated the SFR in the ``{\sfrEXT}'' region using relations that depend on rest-frame UV continuum \citep{salim+07} and on the {\ha} luminosity density \citep{deMellos+24}. As shown below, this should be viewed as upper limits only. 
We obtained an SFR in the ``{\sfrEXT}'' region of $\sim$\,80\,--\,400\,{\msunyr}. This is comparable to the total SFR measured in the system from dust heated by stars of $\sim$\,$140_{130}^{240}$\,{\msunyr} \citep[from SED fitting]{falkendal+19}. 
For comparison, in the ``all'' region (Fig.\,\ref{fig:spectra_regions}), the UV-derived SFR is $\sim$\,$850_{-380}^{+700}$\,{\msunyr}, which is $\sim$\,6 times larger than the IR-based value (when comparing the means). 
Nonetheless, our SFR measurements are rather uncertain. The UV-based measurements, for example, are quite dependent on the relation between the ionized gas and stellar continuum attenuation (see Appendix\,\ref{ap:sfr}). 
Additionally, although we discounted the scattered light (and nebular continuum) contribution in the UV, the percentage of polarization is not based on unresolved measurements \citep[][]{vernet+01}. Therefore, we cannot discard a higher contamination in the extended region. On the other hand, the IR-based value depends on the corrected deblending of the contribution from AGN and stars to the dust-heated IR emission. For {\fourCothree}, only flux upper limits in the rest-frame 75\,--\,110\,{\um} were available, which was the region where the stellar-heated dust emission peaks in the SED model \citep[see Fig.\,A.49 in][]{falkendal+19}. 
Besides that, although IR and UV-based measurements broadly trace similar SFR time scales ($\lesssim$\,200\,Myr), more evolved stars ($\gtrsim$\,100\,--\,200\,Myr) can also contribute to dust-heated IR emission \citep[][]{kennicutt_evans12}. 
This might help explain the discrepancies in the measurements. 
Nonetheless, refined SFR measurements and a deeper discussion about them are beyond the scope of this work.

As noted before, the large-scale radio jet of {\fourCothree} is aligned with the rest-frame UV and optical continua \citep{vanBreugel+98,pentericci+00}. 
Such observations are not uncommon and lead to the hypothesis that jet-driven shocks could trigger local bursts of SF \citep[e.g.,][]{rees89}. This scenario is stronger when co-spatial radio and UV knots are observed \citep[e.g.,][]{labiano+08,duggal+24}, assuming that the latter is tracing newborn stars. 
This could also be the case of {\fourCothree}. However, outside the nuclear region, we only detected radio peaks co-spatial with the rest-frame optical and/or UV peaks in the radio hotspot region (e.g., Fig.\,\ref{fig:fig0}). Nonetheless, this does not rule out that the extended SF could have been triggered in the past by an expanding radio jet, now observed at larger scales ($\sim$\,10\,kpc). Another hypothesis is that the extended SF could be a consequence of previous galaxy interactions, which can cause gas debris to collapse gravitationally \citep[e.g.,][]{barnes_hernquist92}. This could lead to the formation of new stars, which then generates the observed extended UV continuum \citep[e.g.,][]{neff+05}.

\subsection{Outflow kinematics}\label{sec:outflow}

\begin{figure*}
    \includegraphics[width=1\linewidth]{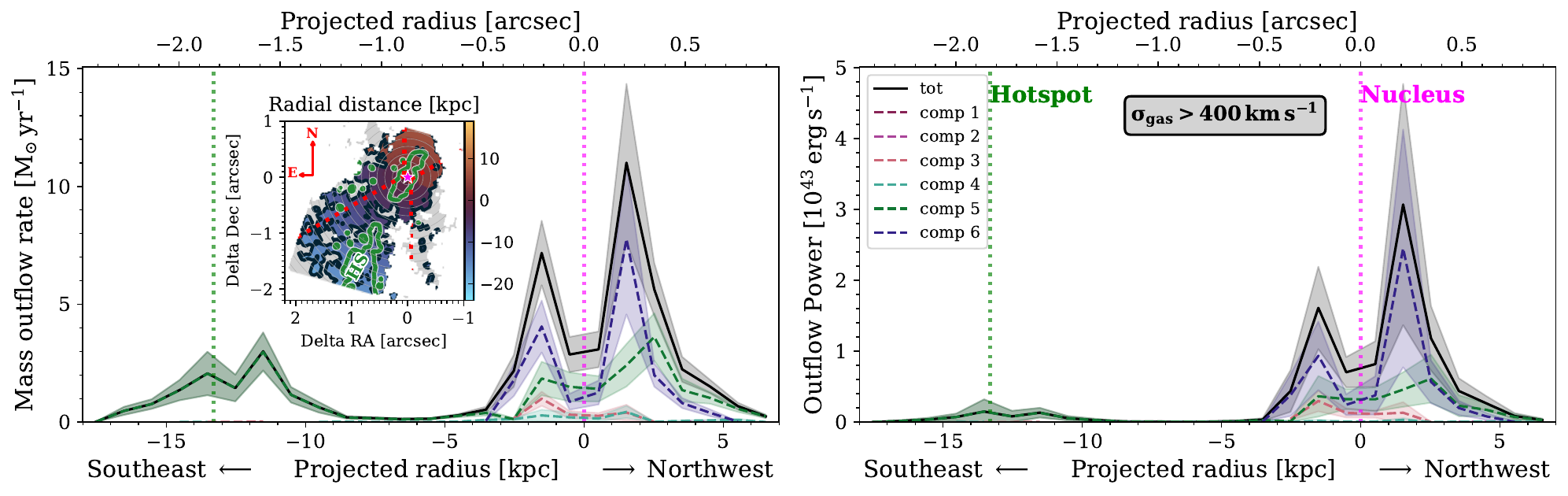}
    \caption{Mass outflow rate (left) and outflow power (right) of {\fourCothree} as a function of the distance from the nucleus. These properties were measured in semi-annuli with 1\,kpc width along the PA\,=\,-30\degree direction, as shown in the inset image. Only individual Gaussian components with $\sigma_{\rm{gas}}$\,>\,400\,{\kms} were considered. The inset map highlights the area where at least one of the components satisfies this condition. The {\colorRadio} contour shows the radio emission, and the dotted red lines mark the biconical ionization region.  
    }
    \label{fig:mdot_edot}
\end{figure*}

In Fig.\,\ref{fig:mdot_edot}, we present the radial profile of the mass outflow rate and the outflow power of {\fourCothree}. The values were derived using {\neSIIcorr} as the electron density, with a velocity threshold cut of $\sigma_{\rm{gas}}$\,>\,400\,{\kms}. We interpret higher values of these quantities as signs of a stronger local impact of the AGN on the ISM. Overall, assuming the same {\nee} for all clouds, we found that $\sim$\,$50$\,\% of the warm ionized gas mass inside $\sim$\,5\,kpc is disturbed ($\sim$\,$30$\,\% for the entire FoV). 

Closer to the nucleus (inside a radius of $r\,{\lesssim}\,3$\,kpc), the ionized gas display the highest disturbances, showing values of {\mdot}\,>\,2\,{\msunyr} and {\edot}\,>\,4\,$\times$\,10$^{42}$\,{\ergs}. In this region, individual Gaussian components can reach velocity dispersions of $\sigma_{\rm{gas}}$\,>\,700\,{\kms} (see Fig.\,\ref{fig:oiii_maps}). This is more pronounced in the 6th component, which shows $\sigma_{\rm{gas}}$ values of up to \,$\sim$\,2000\,{\kms}. It also displays high positive (negative) LoS velocities at the northwest (southeast) of the nucleus, with absolute values of $|v_{\rm{gas}}|$\,$\sim$\,1000\,--\,1500\,{\kms}, indicative of a strong bipolar wind. The warm ionized gas outflow peaks at $\sim$\,2\,kpc to north of the nucleus, reaching values of {\mdotpeak}\,$=$\,$11\,{\pm}\,4$\,{\msunyr} and {\edotpeak}\,$=$\,($3\,{\pm}\,2$)\,{$\times$}\,10$^{43}$\,{\ergs}. 

The {\mdot} and {\edot} values decrease as you move away from the nucleus, but start to increase again at a radius of $\sim$\,10\,kpc to the southeast of it. At $\sim$\,12\,--\,14\,kpc radii, around the radio hotspot, it reach values of {\mdot}\,$\sim$\,2\,{\msunyr} and {\edot}\,$\sim$\,1\,{$\times$}\,10$^{42}$\,{\ergs}. This second peak is associated with the shock front produced at the radio hotspot \citep[see][]{kukreti+26}. 

We can estimate the percentage of the energy released by the AGN that ends up kinematically disturbing the ISM, which is called the kinematic coupling efficiency: $\varepsilon_{\rm{kin}}$\,=\,{\edot}\,/\,{\lagn}. Using the AGN bolometric luminosity of $L_{\rm{AGN}}$\,$=$\,$10^{47.6\,\pm0.1}$\,{\ergs} \citep{falkendal+19} and the peak of the outflow power ({\edotpeak}), we obtained a maximum kinematic efficiency in the warm ($10^4$\,K) ionized gas of $\varepsilon_{\rm{kin}}^{\rm{ion}}$\,$=$\,$8_{-5}^{+7}$\,{$\times$}\,10$^{-3}$\,\% for {\fourCothree}. A same-order  $\varepsilon_{\rm{kin}}^{\rm{ion}}$  value was measured for {4C\,+19.71} \citep{wang+24}, a similar HzRG source, indicating that such efficiencies might be characteristic of such luminous HzRGs.

At first glance, this $\varepsilon_{\rm{kin}}^{\rm{ion}}$ value appears rather low, especially when compared to the thresholds efficiencies needed for the AGN feedback to have an effective impact in the evolution of galaxies, as calculated by theoretical models or used in cosmological simulations \citep[e.g.,][]{zubovas18,nelson+19}. In the literature, one of the lower values for such efficiency thresholds is obtained using the two-stage wind model from \citet{hopkins+10}, which required a $\sim$\,0.5\,\% coupling threshold with the ($10^4$\,--\,$10^6$\,K) ionized gas. In this model, ionized gas winds disturb cold clouds (<\,$10^3$\,K), increasing their surface area, thereby facilitating their subsequent destruction by radiation, which reduces the amount of material that could otherwise form new stars.

However, as discussed by \citet{harrison+18} and \citep{harrison_ramosAlmeida24}, such a comparison is not straightforward, since observations usually measure only the kinetic efficiency of a single gas phase, disregarding the impact on colder and hotter phases. Moreover, {\lagn} is an instantaneous measurement that might have increased or decreased since the launch of the observed outflow. Additionally, we are measuring only the current kinetic power of the gas, ignoring other forms of coupling. This includes the energy used to destroy molecular clouds \citep{garcia-burillo+24,dallagnol+25a}, and losses due to gravitational forces and radiative cooling. As discussed in Sect.\,\ref{sec:cii}, assuming that the {\cii} emission is tracing cold gas (and possibly molecular clouds), the AGN feedback might have depleted part of the cold material along the ionization axis (see Fig.\,\ref{fig:logU_cont}). Such impact is not measured by $\varepsilon_{\rm{kin}}$.

Perhaps a more reasonable comparison can be done with multiphase simulations, which can distinguish the impact on different phases of the gas. \citet{ward+24} tracked the evolution of a spherical wind -- launched at $10^4$\,{\kms} speed -- and its effect on a clumpy medium. They showed that the hot ionized gas phase ({\Tee}\,$\sim$\,10$^{5-8}$\,K, {\nee}\,$\sim$\,0.01\,--\,1\,{\cmcubic}) carries most of the kinetic energy ($\gtrsim$\,90\,\%, for \lagn$\gtrsim$\,$10^{45}$\,{\ergs}), with the effect being more pronounced for higher {\lagn}. Therefore, the warm ionized gas ({\Tee}\,$\sim$\,10$^{4}$\,K, {\nee}\,$\gtrsim$\,$10$\,{\cmcubic}) kinematics -- as measured here for {\fourCothree} -- represent only a fraction of the total outflow power. In follow-up simulations, \citet{almeida+26} showed that the time-averaged  $\varepsilon_{\rm{kin}}^{\rm{ion}}$ for the warm gas decreases for increasing {\lagn}. For {\lagn}\,$\sim$\,$10^{47}$\,{\ergs} (like {\fourCothree}), they found a value of $\sim$\,1.5\,{$\times$}\,10$^{-3}$\,\%, close to our $\varepsilon_{\rm{kin}}^{\rm{ion}}$ value. Note that this is a time-averaged value, which can be about one order of magnitude higher than peak values obtained from radial \edot$(r)$ profiles in the same simulation \citep{ward+24}. Therefore, we expect that $\varepsilon_{\rm{kin}}^{\rm{ion}}$ measurements done mimicking our method should yield <\,10$^{-3}$\,\% values in their simulations.

In these simulations, the warm ionized gas is formed in a post-shock region via radiative cooling of the hot gas wind. This seems in partial agreement with the gas comprised by our 6th Gaussian component (Fig.\,\ref{fig:oiii_maps}), which carries most of the outflow power, and is observed in the region covered by the small-scale radio jet. However, in these simulations the warm ionized gas reach a maximum velocity of $\sim$\,400\,{\kms}, significantly lower than the values observed for the 6th component: \vout\,$\sim$\,1500\,--\,4500\,{\kms}, with {$\sigma$}\,$\gtrsim$\,750\,{\kms}. Another limitation is that these simulations do not track radiative cooling to temperatures $\lesssim$\,$10^3$\,K, meaning that cold clouds are not distinguished from the warm phase.

As originally described by \citet{smail+12} and displayed in Fig.\,\ref{fig:Xray}, {\fourCothree} shows an X-ray emission (rest-frame 0.11\,--\,1.75\,keV) that is spatially correlated with the radio jet. They have interpreted it as inverse Compton emission, originating from far-IR dust-heated photons interacting with relativistic electrons in the jets. Other authors favor a CMB origin for the low-energy photons involved in the process \citep{wu+17,hodges-kluck+21}. 
In the scenario above, where the hot gas carries most of the power \citep{ward+24,almeida+26}, one might associate the extended X-ray with the emission produced by the disturbed hot gas phase at larger scales. However, this extended hot gas is low-density and does not emit efficiently in the X-ray \citep{ward+26}. Therefore, at these redshifts, we should not be able to detect the extended emission from the hot wind with Chandra or other current facilities. For now, observations of X-ray hot winds are restricted to local sources \citep[][]{veilleux+14,lansbury+18,trindade-falcao+26}. 

Regarding the source behind the observed high disturbance close to the nucleus, both AGN photoionization and radio jets might play a role. Both can also play together, with jet-driven shocks enabling the radiation to penetrate further into the ISM \citep{meenakshi+22a}. Additionally, in the proposed scenario of nuclear and extended ongoing SF (see Sect.\,\ref{sec:sfr_ext}), stellar-driven winds should also contribute to the observed turbulence in the medium. The stellar influence should be maximum at the nucleus, where the rest-frame UV peaks, although the effect should be relatively minor compared to the other two mechanisms. AGN radiation-driven winds are likely strong, since AGN photoionization is dominant close to the nucleus, where the ionization parameters reach values of log($U$)\,$\sim$\,$-1.5$. However, we observe two opposing radio spots at $\sim$\,0.2\,--\,0.3{\arcsec} ($\sim$\,2\,kpc) from the nucleus (e.g., see Fig.\,\ref{fig:logU_cont}), inside the most disturbed region. Interestingly, the small-scale jet contours apparently mark the limit of the emission of the 6th Gaussian component, with its kinematic axis being approximately aligned with the jet (see last row of Fig.\,\ref{fig:oiii_maps}, and the dark-purple component in Fig.\,\ref{fig:spec_fits}). 

\citet{nesvadba+17b} estimated the total power associated with the radio jet $L_{\rm{jet}}$\,$=$\,$10^{47.5\,\pm0.3}$\,{\ergs}, using the \citet{cavagnolo+10} relation derived for local massive galaxy clusters. They note that using the \citet{turner_shabala15} relation would increase their value by $\sim$\,0.3\,dex. Therefore, the estimated $L_{\rm{jet}}$ value is close to that of {\lagn}, with both being subject to substantial uncertainties. {\lagn} depends on the correct SED decomposition of the IR emission \citep{falkendal+19}, and on avoiding contamination from nearby sources. The equation used to derive $L_{\rm{jet}}$, which corresponds to the mechanical work done by the jet to inflate X-ray cavities, was derived for more virialized local clusters \citep{cavagnolo+10}. However, we do not expect this to be the case for systems at z\,$\sim$\,3.5. Also, such X-ray cavities are not observed in {\fourCothree} (see Fig.\,\ref{fig:Xray}). Additionally, almost half of the 68\,GHz rest-frame radio luminosity comes from the hotspot in the southeast of {\fourCothree} (see Fig.\,\ref{fig:fig0}), where the jet is likely reaccelerating \citep{vanOjik+95,kukreti+26}, which is not the typical scenario for sources used to derive the above equation. Despite these caveats, we emphasize that both the radio jet and the AGN radiation seem to have enough energy to drive the observed warm ionized gas outflow.

\subsection{Impact of the electron density and other parameters in the outflow measurements}\label{sec:density_outflow}

\begin{figure}
    \includegraphics[width=1\linewidth]{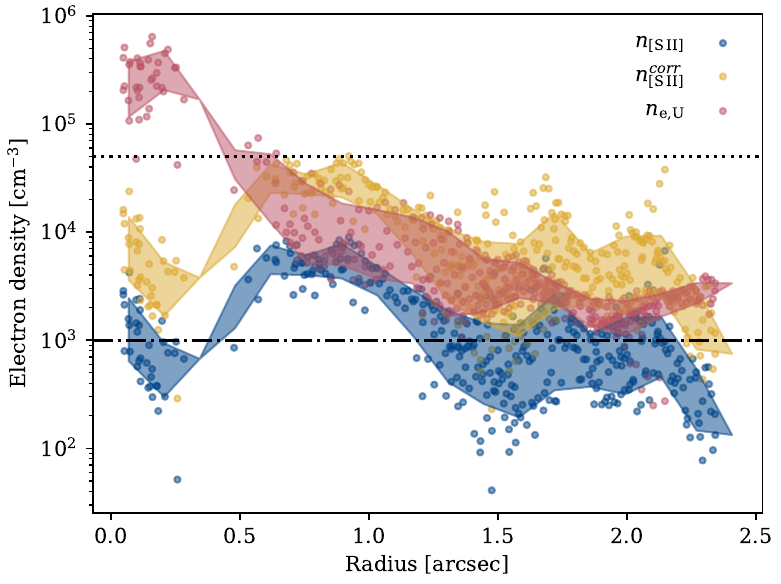}
    \caption{Comparison between different density measurements showing values calculated from the {\sii} lines ({\neSII}, in blue), using \citet{holden+26} correction (\neSIIcorr, in yellow), and from the ionization parameter ({\neU}, in red). Only spaxels with {\neSII} and {\neU} measurements are shown. The black lines correspond to the mean densities of the outflowing ionized gas from \citet{almeida+26} for the entire cloud (dotted line) and for the top 10\% densest cells (dash-dotted line).}
    \label{fig:ne_comparison}
\end{figure}

The electron density tracer used to calculate the outflowing mass {\mout} has an important impact on the outflow kinematics measurements, since ({\mdot}, {\edot})\,$\propto$\,{\mout}\,$\propto$\,1/{\neOut}. In Fig.\,\ref{fig:ne_comparison}, we compare three different density measurements, considering only overlapping spaxels. For the outflow kinematics measured here, we used the simple \citet{holden+26} correction to the {\neSII}, which resulted in a median value of {\neSIIcorr}\,$\sim$\,$(4.7_{3.4}^{10})\,{\times}\,10^3$\,{\cmcubic} (15 and 84\,\% percentiles). Since it is a multiplicative correction, {\neSIIcorr} is higher than {\neSII} by a factor $\sim$\,5.6 everywhere, and inversely for {\mdot} and {\edot}. In Figure\,\ref{fig:mdot_edot_comp}, we display the {\mdot} and {\edot} radial profiles obtained by using these density tracers, and compare them with the results of other tests discussed in this section.

We can compare {\neSII} and {\neSIIcorr} with the electron density obtained from the ionization parameter, which is an independent measurement. The {\neU} measurements vary by $\sim$\,2 orders of magnitude, reaching $\sim$\,$10^5$\,{\cmcubic} at the nucleus. The values decrease for higher radii, approaching {\neSIIcorr} for $r$\,$\gtrsim$\,0.5{\arcsec}, with a good agreement for $r$\,$\gtrsim$\,1.2{\arcsec}. In these regions, the uncorrected {\neSII} values are systematically lower than {\neU}. Similar to our results, \citet{revalski+22} found that {\neSII} tends to underestimate the density at lower radii when compared with density values obtained from detailed photoionization models, with the disagreement being lower at larger distances. However, they did not need to apply corrections to the {\neSII} values to achieve reasonable measurements at the outer radii. This differs from our results, where the best match in these regions is between {\neU} and {\neSIIcorr}. Note, however, that we have used projected radius values to derive {\neU}, which leads to an overestimation of the derived density. This, together with the uncertainties on the ionization parameter and the rate of ionizing photons (see Sect.\,\ref{sec:density}), makes a direct comparison between density methods difficult, and might also partly explain the large discrepancies at $r$\,$\lesssim$\,0.5{\arcsec}. 

Simulations of the AGN wind in a clumpy medium impact from \citet{almeida+26} suggest that the mean density of the outflowing ($\sim$\,$10^4$\,K) ionized gas clouds increases with the AGN luminosity ($\neOut\,{\propto}\,L_{\rm{AGN}}^{0.5}$), with a cap at $10^3$\,{\cmcubic} for {\lagn}\,>\,$10^{46}$\,{\ergs} (the {\fourCothree} case). In the densest regions, however, $\neOut$ keep increasing with the \lagn, reaching $\sim$\,$10^{4.7}$\,{\cmcubic} for the {\fourCothree} luminosities. Our {\neSIIcorr} values are contained between these two values (see Fig.\,\ref{fig:ne_comparison}). We take this as a sign that our {\neSIIcorr} values are representative of the real outflowing gas density, although at the nuclear region, they might be underestimated, as suggested by the {\neU}, and by the fact that the electron density is typically observed to decrease for increasing radii \citep[e.g.,][]{revalski+22}. Using {\neU} would decrease the {\mdotpeak} and {\edotpeak} values by a factor of $\sim$\,8.1 at the nucleus, without significantly affecting the values around the radio hotspot.

Using the same dataset, \citet{roy+26} also quantified the disturbance in the ionized gas in {\fourCothree}. Compared to their measurements, our {\mdotpeak} and {\edotpeak} values are lower by factors of $\sim$\,100\,--\,300 and $\sim$\,30\,--\,200 (considering uncertainties), respectively. Although they used averages as fiducial values (which lower their results by factors of $\sim$\,2\,--\,3), comparing our results with their maximum values is more meaningful. Using VLT/SINFONI observations, \citet{nesvadba+17b} reported even stronger outflow powers in {\fourCothree}, three orders of magnitude higher than our {\edotpeak}. 

Focusing on a comparison between \citet{roy+26} and our {\edotpeak} measurements, most of the difference can be explained by their use of {\neSII} for the electron density. However, even if we use {\neSII} (instead of {\neSIIcorr}), our values would still be lower by factors of $\gtrsim$\,5\,--\,40. Differences in the dust attenuation correction might also be important (see the effect of {\ebv} below). We note that, their method is based on integrating the {\oiii} flux in velocity channels with $|v|$\,>\,500\,{\kms} to obtain the mass, and using {\vout}\,=\,$\sqrt{\Delta v^2+W_{50}^2}$, where $W_{50}$ and $\Delta v$ are the velocity median and its shift from $v_{\rm{rest}}$ in these channels. However, our Gaussian decomposition indicates that, if such deblending is not performed, low-$\sigma_{\rm{gas}}$ material might be misclassified as disturbed gas. This is evident at the north of the nucleus (see Figs.\,\ref{fig:spec_fits} and \ref{fig:oiii_maps}), where the Gaussian components 1\,--\,4 traces warm ionized gas with $v_{\rm{gas}}$\,>\,500\,{\kms}, although having $\sigma$\,<\,400\,{\kms} (interpreted here as non-disturbed clouds). 
We repeat our calculation using a more relaxed threshold of $\sigma_{\rm{gas}}$\,>\,200\,{\kms}, which lead to {\edotpeak} increasing by a factor of $\sim$\,$1.2$. Hence, the impact of selecting a more relaxed $\sigma_{\rm{gas}}$-cut is relatively minor. This happens because, close to the nucleus, components 5 and 6 dominate the outflow kinematics, and the $\sigma_{\rm{gas}}$\,>\,400\,{\kms} cut already selects most of these ``disturbed spaxels.'' 

Finally, we discuss how other parameters can affect our measurements \citep[see also][]{dallagnol+21,riffelRA+23}. Using a different {\vout} definition, such as $\sqrt{\Delta v^2+\sigma^2}$ and $\sqrt{\Delta v^2+3\,\sigma^2}$, would decrease the {\edotpeak} values by factors of $\sim$\,7.6 and 2.5, respectively. Additionally, we ignored projection effects in our calculations. By assuming a fixed outflow inclination (relative to the plane of the sky) of $i_{\rm{out}}$\,=\,20 and 10{\degree}, the outflow power decreases by factors of $\sim$\,3.5 and 14, respectively. On the other side, when testing fixed {\ebv} values of 0.5 and 1\,mag, the {\edotpeak} would increase by $\sim$\,2.1 and 17 factors, respectively. If instead of assuming a constant solar metallicity, we have used the Z map from Fig.\,\ref{fig:Z_Z0} to calculate the mass (see Section\,\ref{sec:mass}), the {\edotpeak} would increase by a factor of $\sim$\,2.6. Overall, by considering the different assumptions and methods discussed in this section, the {\edotpeak} could vary by factors of $\sim$\,1/7.6\,--\,17. This is not a direct constraining of total uncertainty in the measurements, but an assessment of the effect of different assumptions on the results.

\section{Conclusions}\label{sec:conclusions}

We have analyzed the rest-frame optical IFS data of {\fourCothree}, a luminous HzRG at redshift z\,$\sim$\,3.5, based on JWST/NIRSpec observations. By decomposing the emission line profiles into multiple Gaussian components, we were able to isolate the emission from the gas under physical conditions. We found that, inside a 5\,kpc radius, $\sim$\,50\,\% of the warm ($\sim$\,$10^4$\,K) ionized gas mass is disturbed, with the mass outflow rate and power peaking at $\sim$\,2\,kpc. The corresponding kinematic coupling efficiency of $\varepsilon_{\rm{kin}}^{\rm{ion}}$\,$=$\,$8_{-5}^{+7}$\,{$\times$}\,10$^{-3}$\,\% is low but consistent with multiphase simulations that predict that the bulk of the outflow power is carried by the hot ($\sim$\,10$^{5-8}$\,K) ionized gas phase \citep{ward+24,almeida+26}. 

Although {\fourCothree} is one of the most luminous radio sources at this epoch, the ISM ionization is dominated by the AGN photoionization, as highlighted by a well-delineated bipolar region. Such strong radiation, together with the radio jet, may have depleted part of the cold gas, as evidenced by the ``missing'' {\cii}$\lambda$158\,{\um} emission along the ionization axis. In this case, the inner radio emission might be tracing the sides of an expanding radio cocoon, which could be a sign of negative AGN feedback. Stellar winds should also contribute to ISM disturbance, as evidenced by the rest-frame optical and UV continua, which peak at the nucleus. This follows from our assumption that at least part of the continuum emission originates from young stars since scattered light and nebular continuum seem not to be able to fully reproduce the observed emission.
 
The continuum emission also indicates extended SF in the southeast of the nucleus approximately aligned with the radio jet. This could indicate a jet-induced local SF and be interpreted as positive AGN feedback. Another possibility is that the extended continuum traces nearby systems or debris of interactions, which can also induce the collapse of gas clouds, forming new stars in these regions.
Additionally, the gas metallicity distribution suggests a reasonably evolved central galaxy with solar-like metallicities surrounded by metal-poor gas regions, with $Z$ as low as $\sim$\,0.3\,$Z_{\odot}$, possibly associated with nearby companions. However, we emphasize that metallicity measurements carry substantial uncertainties and should be treated with caution. Overall, our results are consistent with the {\fourCothree} system being a proto-cluster, with multiple signs of ongoing galaxy interactions. This scenario might be associated with multiple AGN events, as suggested by the resolved radio jet at different scales of $\sim$\,1 and 10\,kpc. These episodes may then affect the growth of the system associated with the host galaxy positively and/or negatively.

The emerging picture for {\fourCothree} shows that HzRGs at these redshifts can be quite complex systems, where strong radio jets, AGN photoionization, winds, and SF coexist. Isolating such phenomena is possible only due to the high quality of the JWST/NIRSpec data, which highlights the importance of detailed analysis of such observations. In particular, identifying the Gaussian components that trace disturbed warm ionized gas is crucial to avoid overestimating the outflow kinematics. We also reinforce the importance of assessing how different assumptions, such as the parametrization of the outflow velocity and its inclination, can affect the final results. Extra care must be taken with the electron density values. As already discussed in the literature \citep[e.g.,][]{davies+20}, {\sii}-based density measurements should be avoided when inferring the mass of outflowing ionized gas clouds. As an alternative, we applied a simple multiplicative correction to {\neSII} \citep{holden+26}, which yielded {\neOut} values inside the range predicted by multiphase outflow simulations \citep{almeida+26}. Overall, this highlights the importance of considering the influence of different methods and assumptions on the outflow-related measurements, especially when comparing measurements in the literature.

\begin{acknowledgements}

This study was financed in part by the Coordena{\c c}{\~a}o de Aperfei{\c c}oamento de Pessoal de N\'ivel Superior (CAPES-Brasil, 88887.985730/2024-00). 

RR acknowledges support from  Conselho Nacional de Desenvolvimento Cient\'{i}fico e Tecnol\'ogico  (CNPq, Proj. CNPq-445231/2024-6,311223/2020-6, 404238/2021-1, and 310413/2025-7), Funda\c{c}\~ao de amparo \`{a} pesquisa do Rio Grande do Sul (FAPERGS, Proj. 19/1750-2 and 24/2551-0001282-6) and Coordena\c{c}\~ao de Aperfei\c{c}oamento de Pessoal de N\'{i}vel Superior (CAPES, 88881.109987/2025-01).

RAR acknowledges the support from the Conselho Nacional de Desenvolvimento Científico e Tecnológico (CNPq; Projects 303450/2022-3, and 403398/2023-1), the Coordenação de Aperfeiçoamento de Pessoal de Nível Superior (CAPES; Project 88887.894973/2023-00), and Fundação de Amparo à Pesquisa do Estado do Rio Grande do Sul (FAPERGS; Project 25/2551-0002765-9).

CB acknowledges support by a Verbundforschung grant by the German Space Agency (DLR).

This work is based in part on observations made with the NASA/ESA/CSA James Webb Space Telescope. The data were obtained from the Mikulski Archive for Space Telescopes at the Space Telescope Science Institute (STScI), which is operated by the Association of Universities for Research in Astronomy (AURA), Inc., under NASA contract NAS 5-03127 for JWST. These observations are associated with program \# GO-01970. Support for program \# GO-01970 was provided by NASA through a grant from STScI.

The scientific results reported in this article are based on observations made by the Chandra X-ray Observatory and published previously in cited articles.

This paper makes use of the following ALMA data: ADS/JAO.ALMA\#2021.1.00576.S. ALMA is a partnership of ESO (representing its member states), NSF (USA) and NINS (Japan), together with NRC (Canada), NSTC and ASIAA (Taiwan), and KASI (Republic of Korea), in cooperation with the Republic of Chile. The Joint ALMA Observatory is operated by ESO, AUI/NRAO and NAOJ.

Additional software used in this work include: 
\texttt{matplotlib} \citep{matplotlib}, 
\texttt{astropy} \citep{astropy},
\texttt{spectral-cube} \citep{spectralcube},
\texttt{SAOImage DS9} \citep{ds9},
\texttt{QFitsView} \citep{qfitsview},
\texttt{CARTA} \citep{carta2},
\texttt{IFSCUBE} \citep{ifscube2020}.
Paul Tol's\footnote{\url{https://sronpersonalpages.nl/~pault/data/colourschemes.pdf}} and \texttt{cmocean} \citep{cmocean} color schemes were also used.

\end{acknowledgements}

\bibpunct{(}{)}{;}{a}{}{,}
\bibliographystyle{aa_url}
\bibliography{bib}

\begin{appendix}

\section{Ancillary data}\label{ap:data_ancillary}

\subsection{VLT/MUSE rest-frame UV data}

We made use of data taken with the VLT/Multi Unit Spectroscopic Explorer (MUSE, ID:\,60.A-9100(G), \citealt{wang+23}). These observations are part of the MUSE Wide-Field Mode commissioning (FoV of 1{\arcmin}) and cover an observed wavelength range of 4700\,--\,9350\,\AA, with a spatial resolution of 0.63{\arcsec} (FWHM). The rest-frame UV continuum flux distribution was obtained by fitting a 3-degree polynomial, and collapsing the result between the 1029 and 2048\,{\AA} rest wavelengths. For better visualization of the continuum emission (as in Fig.\,\ref{fig:fig0}), we mask the vertical and horizontal artifacts visible in the continuum image and a nearby foreground object (as highlighted in Fig.\,\ref{fig:muse-mask}). The artifacts were visually identified and are associated with the mosaicking of different exposures used to construct a wide-field observation. A {\lyalpha} total flux map was then obtained by integrating the continuum-subtracted cube between 1203 and 1228\,{\AA} rest wavelengths (see contours in Fig.\,\ref{fig:fig0}). 

\subsection{HST rest-frame UV imaging data}
An archival Hubble Space Telescope (HST) image covering the rest-frame UV continuum was also obtained. It corresponds to observations made with an F702W broadband filter \citep[ID: 5932]{pentericci+98}, and covering the 1300\,--\,1800\,{\AA} rest wavelength range. We downloaded the already calibrated data, which is a product of the pipeline Hubble Advanced Products Single Visit Mosaics (HAP-SV). We additionally subtracted the sky background contribution from the image and modified the header to align the position of the rest-frame UV peak emission to the {\fourCothree} nucleus. We also smoothed the original image ($\sim$\,0.2{\arcsec}-FWHM resolution, as measured in field stars) with a 0.3{\arcsec}-FWHM Gaussian kernel, to highlight the weakly detected UV emission \citep[see ][for a non-smoothed version]{pentericci+98}. The resulting image was used for visual comparison with other continuum images.

\subsection{15\,GHz VLA radio data}

To compare the radio jet morphology with maps of different parameters,  we display contours from a VLA radio image centered at 15\,GHz (68.6\,GHz rest-frequency), as in Fig.\,\ref{fig:fig0}. 
The observations were carried out with VLA in A-configuration, and presented in \citet[ID:\,24B-147, PI:\,Carlos De Breuck]{wang+25a}, where the reduction steps are described. The processed image has a synthesized beam with FWHM of 0.13{\arcsec}\,$\times$\,0.15{\arcsec}, and a $\sim$\,4.5\,$\rm{\mu Jy\,beam^{-1}}$ rms noise.

\subsection{ALMA submillimeter and Chandra X-ray data}

Band 8 ALMA observations (2021.1.00576S, PI: Wuji Wang) covering on the {\cii}158{\um} emission line were also used. More specifically, we present contours of flux maps, integrated between three different velocity ranges: 
[$-88$,$\,$99], [99,$\,$324], and [249,$\,$493]\,{\kms}. 
The data reduction is described in \cite{wang+25a}, which resulted in a synthesized beam size of 0.18{\arcsec}\,$\times$\,0.23{\arcsec} and a noise of $\sigma_{\rm{rms}}$ of 92\,{\mujybeam}.

We also present a Chandra 0.5\,--\,8\,keV X-ray image (ID: 12288, PI: Ian Smail) of {\fourCothree}, originally published in \citep{smail+12}. Using the Chandra Interactive Analysis of Observations (CIAO) package, we reprocessed the downloaded dataset with the \texttt{chandra\_repro} command (using default parameters), generating an (level 2) image, showing the distribution of the events. Then, with the CIAO/Sherpa tools in DS9, we smoothed the image using the \texttt{dmimgadapt} algorithm, with the following parameters: \texttt{function}\,=\,gaussian, \texttt{minrad}\,=\,0.5, \texttt{maxrad}\,=\,10, \texttt{numrad}\,=\,30, \texttt{radscale}\,=\,log, and \texttt{counts}\,=\,5. The resulting spatial profiles of nearby point sources in the FoV indicate a resulting spatial resolution of FWHM\,$\sim$\,5{\arcsec}. A comparison between the original and smoothed maps is shown in Fig.\,\ref{fig:Xray}.

\section{Additional details about the fitting procedure}\label{ap:ap_fitting}

\begin{figure}
    \includegraphics[width=1\linewidth]{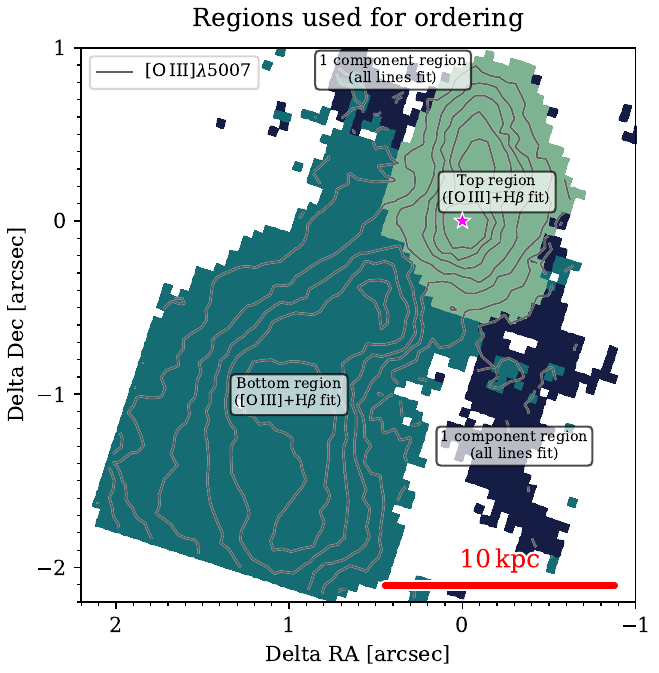}
    \caption{Different regions used to order the Gaussian components of {\fourCothree}. The contours refer to the {\oiii} total flux (see Fig.\,\ref{fig:all_lines}). The top and bottom regions were ordered in independent runs of \texttt{COSMICube}, using the {\oiii} as a reference line. In the remaining regions, where {\oiii} is weak, all emission lines were fitted together with single Gaussian profiles.}
    \label{fig:ordering}
\end{figure}

Here, we describe additional details about the fitting procedure. The numbering (e.g., (2)) refers to the steps described in the main body of the paper (Sect.\,\ref{sec:fitting}). 

For the fit, we used the (inverse of the) uncertainty propagated from the data reduction as the weight of the data. However, for calculating the S/N of the flux density, we used the standard deviation from nearby line-free regions as the noise. More specifically, we measured the standard deviation in multiple 50-pixel-width channels in the continuum-subtracted cube. Then, we modeled the spectral variation of the measured noise with a 4-degree polynomial. 
After this, we modeled the spatial variation along the vertical direction by fitting an 8-degree polynomial to the data normalized by spectral noise variation. We ignored horizontal spatial variation because it was small inside the non-masked regions. The result was a 2D noise map for each emission line, with the same 2D spatial polynomial variation, multiplied by the value of the spectral noise at the observed mean wavelength of the line. 

During the spectral fit, besides fixing kinematics between the emission lines, we applied the following flux constraints between each kinematic Gaussian component:
        {\oiii}$\lambda$5007\,/\,{\oiii}$\lambda$4959\,=\,3,
        {\nii}$\lambda$6583\,/\,{\nii}$\lambda$6548\,=\,3,
        and 
        {\oi}$\lambda$6300\,/\,{\oi}$\lambda$6364\,=\,3. 
        They correspond to the approximate expected ratio between line doublets with the same upper energy level, but different numbers of degeneracies in lower energy levels \citep{osterbrock_ferland06}.     

\paragraph{In step (2).} 
For the initial model of the {\hb} + {\oiii}$\lambda\lambda$4959,5007 emission lines, we restricted the fit to a spectral region covering only this set of lines. For this, we masked the data outside the (4720, 5140)\,{\AA} rest-frame wavelength range.

\paragraph{In step (3).} 
We used the Akaike information criterion (AIC) to decide the proper number of Gaussian components needed to model the observed line profiles. The AIC is defined as \citep{akaike74,schwarz78,sugiura78}: 
\begin{align}
    \rm{AIC} &= -2\ln(\mathcal{L}_{\rm{max}}) + 2k + \frac{2k(k + 1)}{N - k - 1}, \label{eq:aic}
\end{align}
where $N$ is the number of data points and $k$ is the number of parameters of the model, and $\mathcal{L}_{\rm{max}}$ is the maximum likelihood. To evaluate whether we should discard a more complex model (with higher number of Gaussians), a decision criterion is commonly used \citep[e.g.,][]{kass_raftery95}: weak evidence, for $\Delta$AIC < 2; moderate, for 2 < $\Delta$AIC < 6; strong, for 6 < $\Delta$AIC < 10; and decisive, for $\Delta$AIC > 10. We used a threshold of 10. For the maximum likelihood, we used the equation \citep{banks_joyner17}: 
$-2\,\ln(\mathcal{L}_{max}) = N \ln(\chi^2/N) + C$, 
where $\chi^2=\sum_\lambda^N \frac{(F_\lambda-M_\lambda)^2}{\sigma_{F_\lambda}^2}$ is the chi-square, calculated using the flux density of the data ($F_\lambda$), its uncertainty ($\sigma_{F_\lambda}$), and the best-fit model ($M_\lambda$).

\subsection{Ordering of components}\label{ap:ordering}

The result of decomposition of the emission line profiles into multiple Gaussians (Sect.\,\ref{sec:fitting}) is a set of parameter maps for: flux ($F$), LoS velocity ($v$), and velocity dispersion ($\sigma$). One for each  Gaussian component (six in total). Ideally, the parameter ($F$, $v$, and $A$) maps of each component should represent a single physical structure, without abrupt spatial variations (``smooth'' distribution). For example, with component 1 representing gas rotating in a disk, component 2 tracing outflowing gas, and so on. 
Unfortunately, this is usually not the case, as the components must be ordered (or organized) after the fitting procedure.

However, ordering the components is not an easy task. To solve this problem, we used the recently developed software \texttt{COSMICube} \citep{groth25}. This program implements an algorithm that reorganizes the fitted components, aiming to minimize the spatial variance of the parameters ($A$, $v$, and $\sigma$) of each component, both locally and globally. Since all kinematics of all emission lines are tied together, we ran the code on the parameter maps of a single reference line ({\oiii}$\lambda$5007 in our case).
Therefore, the code assumes that individual components are closely separated in their parameter space. The result is a rough ordering, but that can be quite useful to identify physically connected regions.

In step 1, \texttt{COSMICube} roughly organizes the components using a clustering algorithm (\texttt{AgglomerativeClustering}), implemented in the \texttt{scikit-learn} toolkit \citep{scikit-learn}. After this, the algorithm iterates over the spaxels,  reordering Gaussian components in the given spaxel, aiming to minimize the spatial variance of the parameters of all components. Here, the variance of a given parameter is the sum of the absolute differences of the parameter in the spaxel relative to the values of all neighbors: all spaxels inside a square box with a (2\,$\times$\,$d$\,+\,1) side ($d$\,=\,2\,spaxels, in our case). The code starts the iteration from the spaxel with the highest local variance, going to the lowest.

However, testing all possible component permutations (6!\,=\,720, for 6 components) is time-consuming. To circumvent this, the algorithm calculates a cost function \citep[see][for the equations]{groth25}. This function considers together the local and the global variance, after assuming different weights for the variance of each parameter ($A$, $v$, and $\sigma$), and for the global variance. To order the {\fourCothree} Gaussian parameters, we used the following weights ($w$) for the variance of each parameter: ($w_F$, $w_v$, $w_{\sigma}$)\,=\,(0.5, 0.75, 1). Additionally, we used a weight of $w_{\rm{global}}$\,=\,1 for the global variance (relative to the local one). The cost function matrix is minimized using the Jonker-Volgenant algorithm implemented in the \texttt{SciPy} \citep[][\texttt{scipy.optimize.linear\_sum\_assignment}]{scipy20} library. 

\texttt{COSMICube} allows setting the final number of components per spaxel as larger than the number of maximum components fitted. This can be useful, for example, in a situation where the FoV of the observation covers three different clouds, with the emission only superimposing in pairs in each spaxel (never all three together). In this case, a maximum of two Gaussian components would be needed to model each spaxel. Assuming, of course, that a single Gaussian per spaxel models perfectly the emission of individual clouds. 
Allowing a maximum of three components over the entire FoV would better isolate the light of each cloud. 
Nonetheless, for {\fourCothree}, we choose to keep a total of six components (the same number used in the fit).

However, due to the complexity of the emission line profiles in {\fourCothree}, we ran the \texttt{COSMICube} algorithm in two separate regions (see Fig.\,\ref{fig:ordering}). In the ``top'' region, we used a final number of 6 Gaussian components, with a maximum of 5 components being used in the  ``bottom'' region. The algorithm was applied only in the regions where {\oiii} have components with S/N > 3. The third region contains single Gaussian fits, which were obtained by fitting all emission line together. We then joined the three regions together by visually inspecting the ordering solutions, aiming to unite components with overall similar velocity dispersion values. The final result can be visualized in the maps displayed in Fig.\,\ref{fig:oiii_maps}. 

\section{{\ebv} and {\neSII} smoothing maps}\label{ap:smoothing}

The color excess ({\ebv}) map was calculated using smoothed total fluxes of the {\ha} and {\hb} emission lines. We follow these steps: 
(1) calculate the median value of ({\ha}\,/\,{\hb})$^{\rm{mean}}$, using only ``good'' spaxels, with both lines having S/N\,>\,{\SNRratio};  
(2) mask only spaxels with S/N\,<\,{\SNRratiomin} in each flux maps, meaning that all spaxels with S/N\,>\,{\SNRratiomin} will contribute to the smoothing results; 
(3) detect spaxels where the values of the flux ratio are outside the accepted range of values, which is {\ha}\,/\,{\hb}\,>\,2.86 ({\ebv}\,=\,0) in this case; 
(4) in these spaxels, we set {\hb}\,=\,{\ha}\,/\,2.86, forcing the ratio to be within the threshold limit;
(5) smooth both maps with a FWHM\,=\,0.1{\arcsec} Gaussian kernel;
(6) calculate the smoothed ratio ({\ha}\,/\,{\hb})$^{\rm{smoothed}}$, which resulted in the map displayed in Fig.\,\ref{fig:ebv_ne}, where only ``good ratios'' (both lines have S/N\,>\,{\SNRratio}) are shown). 
(7) We then fill ``bad ratio'' spaxels, where {\ha} or {\hb} have S/N\,<\,{\SNRratio}, with the ({\ha}\,/\,{\hb})$^{\rm{mean}}$ value, as calculated in step 1. 
The resulting ({\ha}\,/\,{\hb}) map was then used to calculate the color excess {\ebv}, as described in Sect.\ref{sec:reddening}.

\begin{figure*}
    \includegraphics[width=1\linewidth]{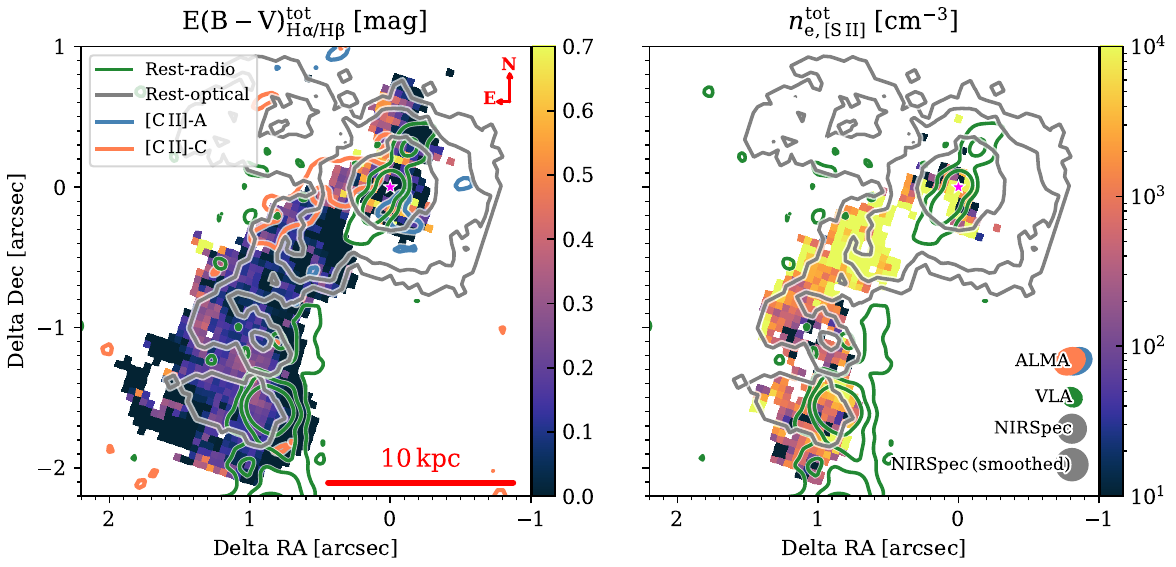}
    \caption{
        Same as Fig.\,\ref{fig:ebv_ne}, but without smoothing the flux maps of the emission lines used to calculate the {\ebv} and {\neSII} maps.
        }
    \label{fig:ebv_ne_raw}
\end{figure*}

We applied the same smoothing method to obtain the electron density of gas, as derived from the {\sii}$\lambda\lambda$6716,31 emission lines ({\neSII}, see Sect.\,\ref{sec:density}). Here, we considered the threshold: {\RneMin}\,<\,{\sii}$\lambda$6716/{\sii}$\lambda$6731\,<\,{\RneMax}, restricting the results to electron densities of {\neMin}\,<\,{\neSII}\,<\,{\neMax}\,{\cmcubic} \citep{sanders+16}. 
Note that, prior to the smoothing, $\sim$\,36\,\% of the spaxels have the {\sii} ratios outside this range, while $\sim$\,31\,\% have {\ha}\,/\,{\hb} were below 2.86. 
We show the smoothed {\neSII} and {\ebv} maps in Fig.\,\ref{fig:ebv_ne}, and the non-smoothed versions in Fig.\,\ref{fig:ebv_ne_raw}. In addition, to obtain the gas metallicity using the {\nii}$\lambda$6543\,/\,{\ha} total flux ratio (see Sect.\,\ref{sec:ionization_parameter} and Fig.\,\ref{fig:Z_Z0}), we also smoothed both fluxes using the above method. In this case, we did not force the {\nii}/\,{\ha} ratios to be inside a given range.

\section{Spectra integrated inside different regions and nebular continuum}\label{ap:spectra_region}

In Fig.\,\ref{fig:spectra_regions}, rest-frame optical continuum integrated spectra from regions. For better visualization, a smoothed version is also displayed, obtained using a Gaussian kernel with 10\,pixels standard deviation. For the last four regions, we also show the rest-frame UV MUSE continuum. In these cases, to match the spatial resolution and projection of the MUSE observation, we smoothed and rebinned the NIRSpec cube before integrating inside the region.
For this, we used the \texttt{reproject\_adaptive} module from the Python package \texttt{reproject} \citep{reproject_20}, with a 0.25{\arcsec}-width Gaussian kernel. 

During the recombination process, when an ion captures an electron, besides the cascade of recombination emission lines (e.g., {\ha} and {\hb}), a continuum spectrum is also generated, known as nebular continuum. Since both events are the result of the same process, we can compare the {\hbeta} ($\gamma_{\hbb}$) and the nebular ($\gamma_{\rm{neb}}$) emissivities, for a given electron temperature ({\Tee}) and density ({\nee}). With the total {\hbeta} flux ($F_{\hbb}$) in a given region, we can obtain the corresponding nebular flux density ($F_{\rm{neb}}$) by integrating side the emitting volume:
\begin{equation}
    \frac{F_{\rm{neb}}}{F_{\hbb}} \propto \frac{\int {\nee} {\npp} \gamma_{\rm{neb}}({\Tee},{\nee})dV}{\int {\nee} {\npp} \gamma_{\hbb}({\Tee},{\nee}) dV} \sim \frac{\gamma_{\rm{neb}}({\Tee},{\nee})}{\gamma_{\hbb}({\Tee},{\nee})},
\end{equation}\label{eq:nebular}
where $n_p$ is the proton density, and the emissivities depend on {\Tee} and {\nee}. For the last steps, we assumed that the emissivities do not vary significantly inside the emitting volume and can be pulled out from the integrals.  

The $\gamma_{\hbb}$ and $\gamma_{\rm{neb}}$ emissivities (in units of $\rm{erg\,s^{-1}\,cm^3}$ and $\rm{erg\,s^{-1}\,cm^3\,\AAA^{-1}}$) were obtained using \texttt{Pyneb} Python package \citep{pyneb_15}, with the internal functions \texttt{get\_continuum} and \texttt{getEmissivity}. As input, we used the mean {\neSII} value inside each region for {\nee}, and a fixed {\Tee}\,=\,$10^4$\,K. The nebular continuum includes free-free, free-bound, and two-photon emission \citep[see][for references of the database used by the package]{pyneb_20}. We also need to inform the ion abundances of He$^{+}$/H$^{+}$ and He$^{++}$/H$^{+}$. We assumed a value of He$^{+}$/H$^{+}$\,=\,0.1, since star-forming galaxies at 1.6\,<\,z\,<\,3.3 display He$^{+}$/H$^{+}$ values in the range of $\sim$\,0.08\,--\,0.11. For log($U$)\,>\,$-2.5$, power law-like ionization sources produce are expected to have abundances of He$^{++}$/He$^{+}$\,$\sim$\,0.1\,--\,1 \citep{dors+22}. To maximize the nebular contribution to the observed spectra, we choose the upper limit, which translates to He$^{++}$/H$^{+}$\,=\,0.1. We note that using He$^{++}$/H$^{+}$\,=\,0.01 would decrease the nebular contribution by factors of $\sim$\,1.1\,--\,1.3 (see Sect.\,\ref{sec:scattering_nebular}). In equation \ref{eq:nebular}, used the dust-corrected $F_{\hbb}$, to obtain the dust-corrected $F_{\rm{neb}}$. We then attenuated the resulting nebular continuum using the mean {\ebv} value within each region (with the same extinction law and assumptions of Sect.\,\ref{sec:reddening}). 

\section{Metallicity and companions}\label{sec:metallicities}

\begin{figure}
    \includegraphics[width=1\linewidth]{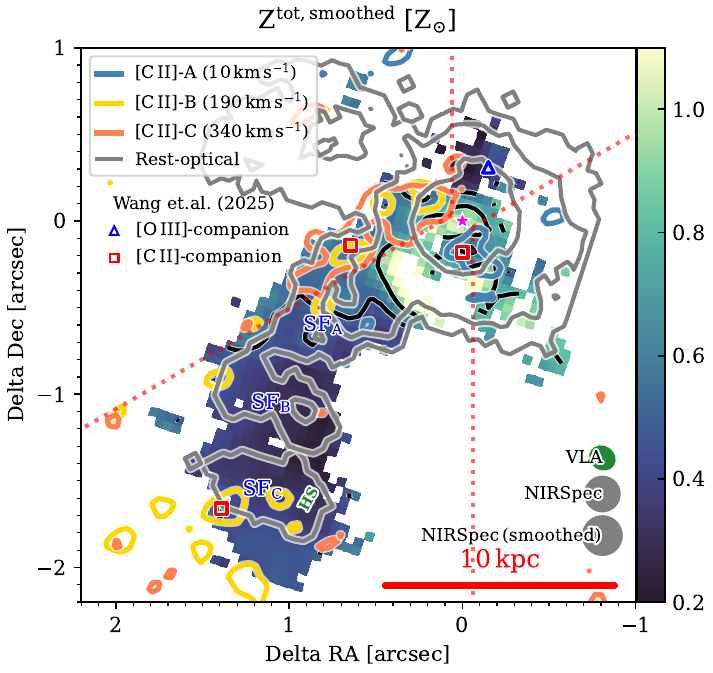}
    \caption{
    Map of the metallicity ($Z$) distribution in the {\fourCothree} system, based on the \citet{carvalho+20} relation, and using the {\nii}$\lambda$6583/{\ha} smoothed total flux ratio. Black contours highlight the 0.5 and 0.8\,$Z_{\odot}$ levels. Overlaid are the contours for the rest-frame optical continuum (in gray, as in Fig.\,\ref{fig:fig0}), and the {\cii}$\lambda$158\,{\um} flux in different channels (\colorCIIa, \colorCIIb, and \colorCIIc, showing only the outer levels of Fig.\,\ref{fig:ebv_ne}). We also draw the locations of the {\oiii} and {\cii} companions \citep{wang+25a}. 
    The red dotted lines delineate the AGN ionizing biconical region (as in Fig.\,\ref{fig:ebv_ne}).}
    \label{fig:Z_Z0}
\end{figure}

Recently, \citet{wang+25a,wang+25b} analyzed Atacama Large Millimeter/submillimeter Array (ALMA) observations covering the [CI](1-0) ($\sim$\,1.75{\arcsec}) and the {\cii}$\lambda$158{\um} ($\sim$\,0.2{\arcsec} resolution) emission lines, which can be used as tracer of cold molecular gas. {\cii} was detected in three different regions, with their inferred masses suggesting a gas-poor cloud origin, ram-pressure stripped from a recent merger \citep{wang+25b}. In the same sense, \citet{villar-martin+07} proposed that recent interactions might have triggered the observed stellar and AGN activity. Motivated by this, we marked the gaseous companion systems proposed by  \citet{wang+25a} in Fig.\,\ref{fig:logU_cont}, calling them {\cii} and {\oiii} companions, with the latter also identified as ``N'' in Figs.\,\ref{fig:spec_fits} and \ref{fig:spectra_regions}. To check how their gas content relates to the AGN host and the extended emission, we plotted their position on the metallicity ($Z$) map in Fig.\,\ref{fig:Z_Z0}. 

For this, we used the {\nii}$\lambda$6583\,/\,{\ha} (N2Ha) total flux ratio as a tracer of $Z$, calculated using smoothed flux maps (see Appendix\,\ref{ap:smoothing}). We applied the equation derived for local Seyfert 2 galaxies from \citet{carvalho+20}: $Z\,/\,Z_{\odot}\,{=}\,4.01^{\log(\rm{N2Ha})}\,{-}\,0.07$, where the solar metallicity ($Z_{\odot}$) corresponds to a solar oxygen abundance of 12 + log(O/H) = 8.69 \citep{asplund+09}. It is important to keep in mind the uncertainties in these $Z$ measurements. We used the \citet{carvalho+20} equation derived ignoring the dependence on $U$, {\nee}, and the power-law spectral index ($\alpha_{\rm{ox}}$) of the ionizing source. For different values of these parameters, their relations return $Z$ values different by up to $\sim$\,0.15$\,\,Z_{\odot}$ and $\sim$\,0.4$\,\,Z_{\odot}$ for low and high metallicites, respectively. Additionally, the results also depend on the assumed relation between the [N/O] and [O/H] abundances, with the abundance relation from \citet{dors+17} being used here. Finally, stellar photoionization should affect local $\rm{N2Ha}$ values, especially where the rest-frame optical (and UV) continuum is relatively stronger (see Sect.\,\ref{sec:continuum}). As discussed in Sect.\,\ref{sec:agn_photo-ionization}, shock-ionization predominates outside the ionization bicone (dashed line in Fig.\,\ref{fig:logU_cont}), and therefore $Z$ values in these regions are more uncertain. We can also expect shock-induced contamination near the radio jet emission, inside the bicone. Nonetheless, the AGN photoionization still dominates the line ratios inside the biconical region, making the relation from  \citet{carvalho+20} suited for constraining the local metallicities in this area. 

In the nucleus, and to the south and southwest of it, we observe a more metal-rich gas with solar-like metallicities ($\sim$\,$1\,Z_{\odot}$). Considering the entire FoV, the $Z$ values are observed inside the $\sim$\,0.2\,\,--\,1.1\,$Z_{\odot}$ range. Overall, together with the evidence of continuum from nearby galaxies (Sect.\,\ref{sec:continuum}), the $Z$ distribution supports the idea of {\fourCothree} being a proto-cluster, with the central galaxy being more metal-rich, and surrounded by low-Z companions. Below, we discuss the metallicity outside the nucleus.

Interestingly, in the areas covered by {\cii}-A and {\cii}-C companions (see contours in the southeast and southwest of the nucleus), we find lower values of 0.5\,$\lesssim$\,$Z\,/\,Z_{\odot}$\,$\lesssim$\,0.8, with {\cii}-B showing even lower values. In general, relative to the surrounding gas, the {\cii} companions are located in regions with lower metallicities, with higher attenuation (Fig.\,\ref{fig:ebv_ne}) and ionization parameter (Fig.\,\ref{fig:logU_cont}). Furthermore, they are observed close to or inside the rest-frame optical continuum, but without signs of local continuum enhancement. Therefore, they might not be tracing the exact location of galaxy companions, but gas stripped from them. They can also be associated with gas that survived the AGN negative feedback, maybe due to dust shielding, since they are located at the edges of the biconical region (Fig.\,{\ref{fig:logU_cont}}).

However, the {\oiii} (or ``N'') companion, situated at $\sim$\,0.35{\arcsec} ($\sim$\,2.5\,kpc projected distance) to the north of the nucleus, seems to be actually tracing a nearby galaxy. In favor of this, we found considerably lower metallicities (compared to the nucleus) of 0.3\,$\lesssim$\,$Z\,/\,Z_{\odot}$\,$\lesssim$\,0.5. Similar values are found in the SF$_{\rm{B}}$ and SF$_{\rm{C}}$ regions, where we have found evidence of extended SF. Moreover, unlike the rest-frame optical continuum spectrum at the nucleus (``AGN''), the ``N'' spectrum continues to rise to shorter wavelengths (see Fig.\,\ref{fig:spectra_regions}). This could be a sign of ``N'' and ``AGN'' having different stellar populations. However, other effects, including a different dust attenuation and/or contribution from nuclear scattered light, might also affect the spectral shape. In addition, unlike the {\cii} companions, there is a small elongation in the rest-frame optical continuum at the ``N'' location (better seen in Fig.\,\ref{fig:all_lines}).  

A different scenario was proposed by \citet{roy+26} as an explanation for the ``N'' region. They favor shock-induced disturbances, due to the jet propagating in a nonuniform ISM in this region. Their conclusions are partly based on the observed high velocity shifts and total width of {\oiii} profile ($w_{80}$\,$\sim$\,1500\,{\kms}), together with the presence of radio jet emission at the ``N'' region. Supporting this, based on our emission line modeling, we found more than one Gaussian component tracing disturbed gas ($\sigma$\,>\,400{\kms}) in this region. We do not discard that the jet plays an important role in disturbing that gas locally, since there is a radio peak at this spot. However, we also detected more than one component with $\sigma$\,<\,$200${\kms} (see the ``N'' spectrum in Fig\,\ref{fig:spec_fits}), indicating the presence of less disturbed gas kinematics, which reinforces the presence of a minor companion. We note, however, that the hypothesis of ``N'' being a companion only holds if we are observing the first flyby of this minor galaxy, since multiple encounters would reduce the relative velocity between them, which can reach $\sim$\,2000\,{\kms} for the Gaussian component 2. High relative velocities $\sim$\,1000\,--\,2000\,{\kms} between interacting systems have been observed in the z\,$\sim$\,2 merging proto-cluster known as Spiderweb \citep{kuiper+11}, although at higher projected distances of $\sim$\,20\,--\,50\,kpc. Another example is the z\,$\sim$\,3 Step proto-cluster which include four-to-seven galaxies, spanning 800\,{\kms} relative LoS velocities, separated by 20\,--\,70\,kpc \citep{bertemes+26}.

In the emission perpendicular to the ionization axis, along the southwest-northeast direction, the few spaxels with  S/N\,>\,3 show $Z$\,$\sim$\,0.5\,--\,0.8\,$Z_{\odot}$, also lower in comparison with the nucleus. The ionized gas in this region, as traced by {\oii} and {\ha}, is observed up to $\sim$\,2{\arcsec} ($\sim$\,15\,kpc) to the southeast (see Fig.\,\ref{fig:all_lines}). This is also a region with extended {\lyalpha} (see Fig.\,\ref{fig:fig0}). Based on the {\lyalpha} kinematics, \citet{vanOjik+95} proposed that we might be observing an inflow of gas from larger scales. We additionally proposed that we are observing an ongoing interaction, where the companion ``collided'' with the {\fourCothree} host coming from the southwest of the nucleus. In this case, the rest-frame optical ``artifact'' of Fig.\,\ref{fig:logU_cont} could actually be debris of this interaction. The lower relative metallicity also supports this hypothesis. Note, however, that shocks are the main driver of ionization in this region \citep{kukreti+26}. Therefore, the \citet{carvalho+20} relation (used to derive $Z$) is probably not suited for determining the local metallicity there. 

Finally, on a larger scale (spanning $\sim$\,90\,kpc), there is a long arc-shaped emission in the VLT/MUSE rest-frame UV continuum, from the southeast to the northwest (bottom left to top right of Fig.\,\ref{fig:fig0}). It is tempting to assign this ``structure'' to physically connected sources. At least for the northwest continuum blob, we have further evidence of it being part of the {\fourCothree} system. This is based on the VLT/SINFONI observations presented in \citet{nesvadba+17a}, covering a larger FoV than JWST/NIRSpec. Their Figure\,A.5 shows an {\oiii} emission at $\sim$\,2{\arcsec} to the northwest of the nucleus, and $\sim$\,1{\arcsec} to the east of the northernmost radio emission. This emission coincides with part of the northeast UV continuum, although the local UV emission peaks $\sim$\,1{\arcsec} to the north of it. In this region, their {\oiii} maps show $\sim$\,$-250$ to 0\,{\kms} velocity shifts, which reinforce the hypothesis of the continuum being associated with a companion galaxy. Additionally, archival near-IR Spitzer/IRAC observations \citep[$\sim$\,0.8\,--\,1.8\,{\um} rest-wavelength]{deBreuck+10} show resolved emission in the region where we find the {\oiii} and UV continuum blobs. For the southeast UV emission (bottom left of Fig.\,\ref{fig:fig0}), there is no evidence of it being associated with {\fourCothree}.

\section{Star formation rate}\label{ap:sfr}

We first estimated the SFR using the MUSE rest-UV spectra, integrated in the ``all'' and ``{\sfrEXT}'' regions (see Fig.\,\ref{fig:spectra_regions}). We used the relation from \citet{salim+07}, which is calculated from the dust-corrected UV luminosity density  ($L_{\rm{UV}}$), measured in the 1300\,--\,1800\,{\AA} rest-frame range. 
To calculate $L_{\rm{UV}}$, we corrected the rest-frame continuum by the dust attenuation using the \citet{calzetti+00} law. For this, we assume a fixed relation between the color excess in the stellar continuum ({\ebv}$_{\rm{star}}$) and in the ionized gas ({\ebv}$_{\rm{gas}}$) of {\ebv}$_{\rm{star}}$\,$=$\,{\ebv}$_{\rm{gas}}$\,+\,0.44\,mag \citep{calzetti97}. For {\ebv}$_{\rm{gas}}$, we used the mean value (inside each region) obtained from the {\ha}\,/{\hb} flux ratio (see Sect.\,\ref{sec:reddening} and Fig.\,\ref{fig:ebv_ne}). 
Before correcting for dust attenuation and calculating $L_{\rm{UV}}$, we discounted the contribution (in the observed continuum) from the scattered light \citep[fixed 11\,\% fraction]{vernet+01}, and from the nebular continuum (see Appendix\,\ref{ap:spectra_region}). 
The resulting SFR$_{\rm{UV}}$ values in the  ``{\sfrEXT}'' and ``all''  regions were $\sim$\,$220_{-100}^{+180}$\,{\msunyr} and $\sim$\,$850_{-380}^{+700}$\,{\msunyr}, respectively. 
Note that these measurements are highly dependent on the dust correction. By using {\ebv}$_{\rm{gas}}$ instead of {\ebv}$_{\rm{star}}$ to correct for the dust attenuation, the SFR$_{\rm{UV}}$ values decrease by a factor of $\sim$\,40. 
Additionally, although we have measured $L_{\rm{UV}}$ using the masked MUSE (see Fig.\,\ref{fig:muse-mask}), the mosaicking artifacts might still contaminate the results. 

In the ``{\sfrEXT}'' region, we also estimated the SFR from the total dust-corrected {\ha} luminosity. Since this region is dominated by AGN photoionization, we used the relation from \citet{deMellos+24}, derived for local sources using only regions with AGN-like ionization. The relation is obtained by comparing the surface density of the {\ha} luminosity ($\sum L_{\rm{\haa,AGN}}$) and the SFR ($\sum\rm{SFR}_{\rm{\haa,AGN}}$), with the latter derived by stellar population synthesis in \citet{riffel+21}. Therefore, we follow these steps: (1) calculate the $\sum L_{\rm{\haa,AGN}}$ inside the ``{\sfrEXT}'' region; (2) obtain $\sum\rm{SFR}_{\rm{\haa,AGN}}$ from the \citet{deMellos+24} relation; multiply the result by the area of the region. This resulted in an SFR$_{\rm{\haa,AGN}}$\,$\sim$\,$150_{-70}^{+120}$\,{\msunyr} inside the {\sfrEXT} region. The reported 1-$\sigma$ errors of SFR$_{\rm{\haa,AGN}}$ and SFR$_{\rm{UV}}$ are dominated by the uncertainties of the relations used in the calculations.

\section{Additional figures}\label{ap:figures}

\begin{figure}
    \includegraphics[width=1\linewidth]{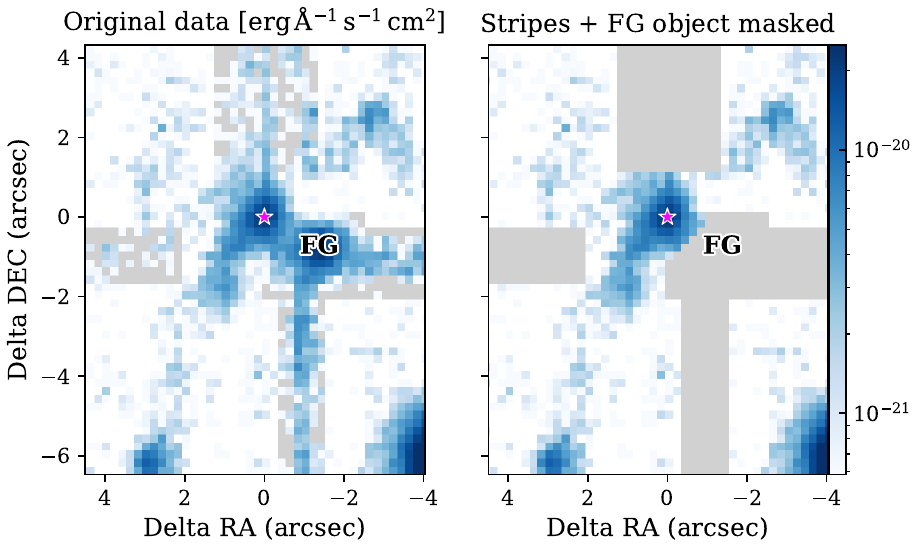}
    \caption{MUSE rest-frame UV continuum maps of {\fourCothree}, highlighting the regions that were manually masked (gray regions). Left: original reduced data. Right: Image, after the nearby foreground object and horizontal and vertical artifacts (associated with the mosaicking of different exposures) were masked.}
    \label{fig:muse-mask}
\end{figure}

\begin{figure*}
    \includegraphics[width=1\linewidth]{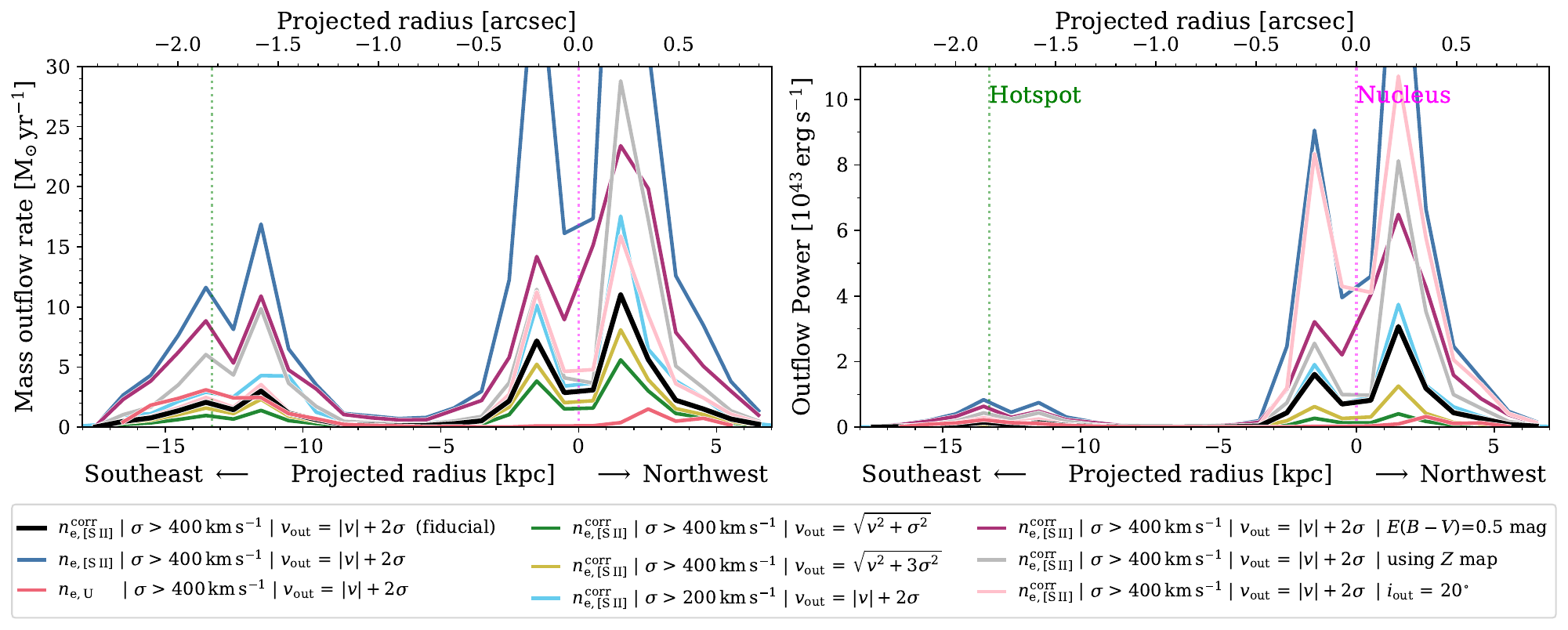}
    \caption{Same as Fig.\,\ref{fig:mdot_edot}, but showing the effect of different assumptions on the mass outflow rate (left) and outflow power (right) radial profiles of {\fourCothree} (see discussion in Sect.\,\ref{sec:density_outflow}). The black lines correspond to our fiducial {\mdot} and {\edot}. For each colored line, the legend shows which electron density tracer, outflow velocity parametrization, and $\sigma_{\rm{gas}}$ threshold were used in the calculation. The last lines show the effect of assuming fixed values of {\ebv}\,=\,0.5\,{mag} and {\iout}\,=\,20{\degree}. It also shows the effect of using the estimated metallicity map ($Z$, see Fig.\,\ref{fig:Z_Z0}) to calculate the gas mass (instead of assuming a solar value). 
    Note that the {\mdot} and {\edot} values of the {\neSII} test (second line) differ from the fiducial radial profiles by a fixed factor of $\sim$\,5.6.}
    \label{fig:mdot_edot_comp}
\end{figure*}

\begin{figure}
    \includegraphics[width=1\linewidth]{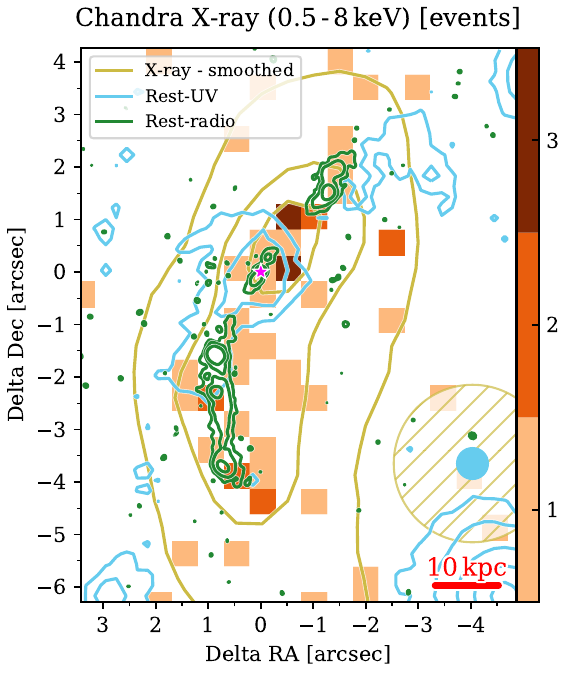}
    \caption{Reconstructed image of the X-ray Chandra 0.5\,--\,8\,keV detected events distribution. The {\colorXray} contours correspond to the smoothed image (FWHM\,$\sim$\,5{\arcsec}), showing levels at [0.11, 0.22, 0.43]\,counts. The {\colorRadio} and {\colorUV} contours correspond to the rest-frame radio and UV continua, respectively (as in Fig.\,\ref{fig:fig0}).}
    \label{fig:Xray}
\end{figure}

Here, we display extra figures that are mentioned in the main part of the text. 
Figure \ref{fig:muse-mask} highlights the masked regions in the VLT/MUSE rest-frame UV continuum image, which were used in Fig.\,\ref{fig:fig0}. The same spatial regions were masked in the MUSE cube before obtaining the spectra shown in Fig.\,\ref{fig:spectra_regions}. A comparison between {\mdot} an {\edot} radial profiles obtained using different assumption are shown in Figure\,\ref{fig:mdot_edot_comp}.
We also display the reconstructed X-ray map and its smoothed contours in Fig.\,\ref{fig:Xray}, which was used in the discussion of Sect.\ref{sec:outflow}.

\end{appendix}

\end{document}